\documentclass[11pt]{article}
\usepackage{amsmath,amsfonts,amsthm,amssymb,latexsym}
\usepackage{bm,bbold}
\usepackage{comment}
\usepackage{graphicx}
\usepackage[margin=1in,letterpaper]{geometry}
\usepackage{indentfirst}
\usepackage{natbib}
\usepackage{soul}
\usepackage{enumitem}

\usepackage[ruled]{algorithm2e}   %
\SetKwInput{KwInput}{Input} 
\SetKwInput{KwOutput}{Output}

\newcommand{\lp}[1]{\left(#1\right)}

\newcommand{\lB}[1]{\left\{#1\right\}}

\newcommand{\lV}[2][]{\left\|#2\right\|_{#1}}

\newcommand{\indicator}[1]{\mathbb{1}_{\lB{#1}}}
\newcommand{\tran}{^\mathsf{T}}

\newcommand{\ietext}{\text{i.e. }}
\newcommand{\egtext}{\text{\mbox{e.g. }}}

\newcommand{\R}{\mathbb{R}}

\theoremstyle{remark}

\theoremstyle{definition}

\theoremstyle{definition}

\theoremstyle{definition}

\theoremstyle{definition}

\newcommand{\bmw}{\bm{w}}
\newcommand{\bmx}{\bm{x}}

\newcommand{\bmS}{\bm{S}}

\newcommand{\bmtheta}{\bm{\theta}}

\newcommand{\bmmu}{\bm{\mu}}

\newcommand{\bmSigma}{\bm{\Sigma}}

\newcommand{\bfone}{\mathbf{1}}

\usepackage{subcaption}
\usepackage{caption}
\usepackage{booktabs}
\usepackage{multirow}  %
\usepackage{afterpage}
\usepackage{changepage}
\usepackage{float}
\usepackage{setspace}
\usepackage{csquotes}
\usepackage{makecell}

\newcount\Comments  %
\usepackage{color}
\definecolor{darkgreen}{rgb}{0,0.5,0}
\definecolor{purple}{rgb}{1,0,1}

\usepackage{hyperref}
\usepackage{xcolor}
\hypersetup{
	colorlinks,
	linkcolor={blue!50!black},
	citecolor={blue!50!black},
	urlcolor={blue!80!black}
}
\usepackage{tablefootnote}

\title{Dynamic Asset Allocation with Asset-Specific Regime Forecasts}   %

\author{
Yizhan Shu\footnote
{Department of Operations Research and Financial Engineering, Princeton University, Princeton, USA} \and
Chenyu Yu\footnotemark[1]\, \footnote{The first two authors contribute equally to the article.}  \and
John M. Mulvey\footnotemark[1]  
\,\footnote{
Yizhan Shu  (corresponding author), \href{yizhans@princeton.edu}{yizhans@princeton.edu}, 
Chenyu Yu, \href{chenyu@princeton.edu}{chenyu@princeton.edu}, 
John M. Mulvey, \href{mulvey@princeton.edu}{mulvey@princeton.edu} 
}
}

\date{\today}

\begin{document}

\maketitle

\begin{abstract}

This article introduces a novel hybrid regime identification-forecasting framework designed to enhance multi-asset portfolio construction by integrating asset-specific regime forecasts. 
Unlike traditional approaches that focus on broad economic regimes affecting the entire asset universe, our framework leverages both unsupervised and supervised learning to generate tailored regime forecasts for individual assets.
Initially, we use the \emph{statistical jump model}, a robust unsupervised regime identification model, to derive regime labels for historical periods, classifying them into bullish or bearish states based on  features extracted from an asset return series.  
Following this, a supervised gradient-boosted decision tree classifier is trained to predict these regimes using a combination of asset-specific return features and cross-asset macro-features. 
We apply this framework individually to each asset in our universe. 
Subsequently, return and risk forecasts which incorporate these regime predictions are input into Markowitz mean-variance optimization to determine optimal asset allocation weights. 
We demonstrate the efficacy of our approach through an empirical study on a multi-asset portfolio comprising twelve risky assets, including global equity, bond, real estate, and commodity indexes spanning from 1991 to 2023.
The results consistently show outperformance across various portfolio models, including minimum-variance, mean-variance, and naive-diversified portfolios, highlighting the advantages of integrating asset-specific regime forecasts into dynamic asset allocation.

\end{abstract}


\newpage

\section{Introduction}

In his seminal 1952 article, Harry Markowitz introduced a groundbreaking two-stage approach to portfolio selection, comprising forecasting and optimization \citep{markowitz52}. %
In the second stage of optimization, he advocated the use of the ``expected returns -- variance of returns''  rule, aiming to maximize expected portfolio returns while minimizing risk.     %
Markowitz’s identification of variance as a critical measure of portfolio risk established a quantitative foundation for understanding diversification, as an ``observed and sensible'' investment behavior, through the covariance among securities.
Academically, his portfolio theory spurred fundamental developments in modern financial economics and operations research, including the Capital Asset Pricing Model (CAPM) \citep{sharpe64} and Quadratic Programming (QP) algorithms \citep{markowitz56, wolfe59}. 
Industrially, Markowitz’s pragmatic approach found significant resonance, leading to the widespread adoption of his two-stage framework by quantitative asset management firms and hedge funds.

The challenge of achieving accurate forecasts in the first stage of Markowitz’s framework is well-recognized among practitioners; indeed, with low-quality forecasts, the optimization stage can become an ``error maximizer'' \citep{michaud1989}. 
Markowitz’s original article acknowledged this challenge, suggesting the use of ``statistical techniques and the judgment of practical men'' applied to historical data as a general solution. 
He also expressed confidence that ``better methods, which take into account more information, can be found''.
This article explores how one such source of ``more information'' -- financial market regimes -- can enhance multi-asset portfolio construction. 
In our approach, market regimes identified by an unsupervised algorithm serve as the forecasting targets, which exhibit an enhanced signal-to-noise ratio compared to direct forecasts of asset returns. 
This enhancement facilitates more accurate prediction by the subsequent forecasting algorithm. 
By strategically applying forecasting skills to these identified market regimes, we seek to improve the robustness of the forecasting stage and thus enhance the practical effectiveness of the Markowitz portfolio theory.

The cyclical regime-switching behavior of asset returns is well-documented across various asset classes, including equities \citep{hardy2001}, fixed income \citep{gray1996}, commodities \citep{Choi2010}, currencies \citep{REUS2016}, and cryptocurrencies \citep{Ardia2019}. 
This behavior manifests itself in patterns such as bullish or bearish markets \citep{Pagan2003}, periods of high or low volatility \citep{Schwert1989}, and risk on or off episodes \citep{bock2015}. 
A specific regime denotes extended, consecutive periods that display homogeneous market dynamics regarding returns, volatility, and sentiment, among other factors, while a regime shift signifies an abrupt yet persistent change in market behavior. 
Regime-switching models are highly valued for their interpretability, as the regimes they identify often correlate with real-world events like different phases of the business cycle \citep{hamilton1989} or shifts in monetary policy \citep{ang2002monetary}. 
These models not only align with changing fundamentals that can often be interpreted only ex-post, but also facilitate ex-ante real-time forecasting \citep{ang2012}.
Notably, \citet{ang2004} was among the first to empirically demonstrate the benefits of incorporating regimes into asset allocation in an out-of-sample setting.

Previous studies on regime-based  asset allocation typically focus on identifying or predicting a limited number (usually 2--4) of economic regimes, such as expansion and recession periods, and subsequently employ regime-dependent strategies. 
These economic regimes are often inferred from macroeconomic indicators \citep{kim2023, Glissmann2024}, or derived from a major equity index using Markov-switching models \citep{nystrup2015asset}. 
Regime-dependent actions might involve tilting weights towards favorable assets under each regime \citep{Nystrup2018}, employing dynamic risk budgets \citep{james19}, or integrating regime-dependent return and covariance forecasts into portfolio optimization \citep{Uysal2021}. 
This approach presumes that a few broad economic regimes can explain the variation in time-varying asset behaviors uniformly across the entire asset universe -- for instance, stocks typically underperform during recessions, while bonds may outperform.
While the methodology’s strength lies in its ability to make asset performance interpretable in terms of the macro-environment, it may be overly optimistic to believe that such a simplified set of regime concepts can adequately capture the complex behaviors of a diverse range of assets.

In this article, we introduce a hybrid regime identification-forecasting framework that is applied to every asset in our universe to derive asset-specific regime forecasts\footnote{
Our approach draws inspiration from \citet{Bosancic2024}, who generates regime forecasts for a collection of individual factor portfolios. 
Additionally, \citet{ang2004} also hinted ``\ldots we used only one regime variable, but it would be interesting to accommodate \emph{country-specific regimes} \ldots''.
}.
Initially, we utilize the  \emph{statistical jump model} (JM)\footnote{
The JMs discussed in this article are not related to \emph{jump-diffusion models}, commonly used in stochastic finance. 
Throughout this article, we will use the terms statistical jump model, jump model, and JM interchangeably.
} to classify historical periods into bullish and bearish  regimes, based on  features extracted from the asset return series.  %
These labels are then shifted forward by one day to serve as the prediction targets for a gradient-boosted decision tree classifier in the following step, using an expanded set of features. 
This two-step ``unsupervised-supervised'' framework enables the regime identification algorithm to generate regime labels that facilitate accurate prediction in the subsequent  forecasting step.   %
A critical aspect of our methodology is the tuning of the jump penalty, a key hyperparameter in JMs that moderates the signal-to-noise ratio of the identified labels. 
We determine each asset’s optimal jump penalty through a time-series cross-validation approach that seeks to maximize the validation performance of a simple regime-switching strategy applied to the asset. 
This strategy, initially introduced by \citet{bulla2011asset}, provides an effective evaluation of the financial implication of the statistical accuracy of regime forecasts for a specific asset.
Our methodological shift from broad economic conditions to specific market dynamics enhances the predictability and relevance of our regime forecasts.

Following the derivation of regime forecasts through our hybrid framework, the subsequent step involves translating these predictions into asset allocation weights. 
We illustrate this process by integrating the regime forecasts into three distinct portfolio construction models: minimum-variance (\mbox{MinVar}), mean-variance (MV), and equally-weighted (EW) portfolios.
The MinVar and MV portfolios utilize a formal Markowitz mean-variance optimization, incorporating appropriate constraints and objective terms. 
The integration of asset-specific regime predictions into the Markowitz optimization is intuitive and straightforward, facilitating dynamic asset allocation that optimizes the risk-return trade-off.

In our empirical study, we apply our methodology to a multi-asset portfolio comprising twelve diverse assets, including global equity, bond, real estate, and commodity indexes from 1991 to 2023. 
Initially, we visually showcase our method's ability to capture the distinct market regimes for various assets through illustrative figures of regime forecasts. 
Furthermore, we present the financial implication of the accuracy of our regime forecasts for all individual assets, and demonstrate consistent outperformance across three portfolio models when integrating our regime forecasts.

The remainder of the article is organized as follows. 
Section \ref{sec:asset universe} details the data used in our empirical study, providing descriptions along with descriptive plots. 
In Sections \ref{sec:regime} and \ref{sec:portfolio}, we delve into our methodology and discuss related literature, with Section \ref{sec:regime} focusing specifically on the hybrid regime identification-forecasting framework, and Section \ref{sec:portfolio} on the portfolio construction methods.
For a comprehensive summary of all methodological steps involved in our study, refer to Section \ref{subsec:method summary}.
Section \ref{sec:results} presents and discusses our empirical results. 
Finally, Section \ref{sec:conclusion}  concludes.

\section{Asset Universe} \label{sec:asset universe}  %

In this article, we consider an asset universe consisting of twelve risky assets and one risk-free asset. 
The risky assets include US Large-Cap, Mid-Cap, and Small-Cap Equity (abbreviated as \texttt{LargeCap}, \texttt{MidCap}, \texttt{SmallCap}), Developed Markets ex-US Equity (\texttt{EAFE}), Emerging Markets Equity (\texttt{EM}), US Aggregate Bonds (\texttt{AggBond}), US Long Treasury Bonds (\texttt{Treasury}), US Liquid High Yield Bonds (\texttt{HighYield}), US Investment-Grade Corporate Bonds (\texttt{Corporate}), US Real Estate (\texttt{REIT}), Commodities (\texttt{Commodity}), and Gold (\texttt{Gold})\footnote{
We will use these abbreviations in monospaced font to refer to the asset classes throughout this article.
}, as detailed in Table \ref{tab:assets}. 
We sourced daily total return indexes denominated in USD for these assets from 1991 to 2023 via the Bloomberg Terminal. 
Each asset class is represented by a major index tracked by an actively traded exchange-traded fund (ETF) to ensure practical applicability. 
For instance, \texttt{LargeCap} corresponds to the \mbox{S\&P 500} index and \texttt{AggBond} to the Bloomberg US Aggregate Bond Index. 
A comprehensive description of the indexes and their corresponding ETFs is provided in \mbox{Appendix \ref{sec: data details}}. 
Additionally, we use the US Treasury three-month constant maturity yield as a proxy for the risk-free rate, sourced from the Federal Reserve Economic Data (FRED) database\footnote{Federal Reserve Economic Data (FRED), Federal Reserve Bank of St. Louis, \url{https://fred.stlouisfed.org}.}.
Overall, our portfolio maintains a long-only, 100\% allocation across these twelve risky assets and one risk-free asset.

\begin{table}[htbp]   
\centering
\begin{tabular}{cc}
\toprule
\textbf{Category} & \textbf{Asset (Abbreviation)}  \\ 
\midrule
\multirow{3}{*}{US Equity} & US Large-Cap Equity (\texttt{LargeCap})  \\
                            & US Mid-Cap Equity (\texttt{MidCap}) \\
                            & US Small-Cap Equity (\texttt{SmallCap}) \\ 
\midrule
\multirow{2}{*}{World Equity} & Developed Markets ex-US Equity (\texttt{EAFE})  \\
                                & Emerging Markets Equity (\texttt{EM})  \\ 
\midrule
\multirow{4}{*}{US Bonds} & US Aggregate Bonds (\texttt{AggBond}) \\
                        & US Long Treasury Bonds (\texttt{Treasury}) \\
                        & US High Yield Bonds (\texttt{HighYield})\\
                         & US Corporate Bonds (\texttt{Corporate})  \\ 
\midrule
  US Real Estate                   & US Real Estate (\texttt{REIT})  \\ 
\midrule
\multirow{2}{*}{Commodity} & Commodities (\texttt{Commodity}) \\
                            & Gold (\texttt{Gold})\\ 
\midrule
Risk-Free & US Treasury 3-Month Yield  \\
\bottomrule
\end{tabular}
\caption{Overview of the asset universe employed in this article, comprising twelve risky assets and one risk-free asset.}
\label{tab:assets}
\end{table}

\begin{figure}[htbp]
    \centering
    \includegraphics[width=\textwidth]{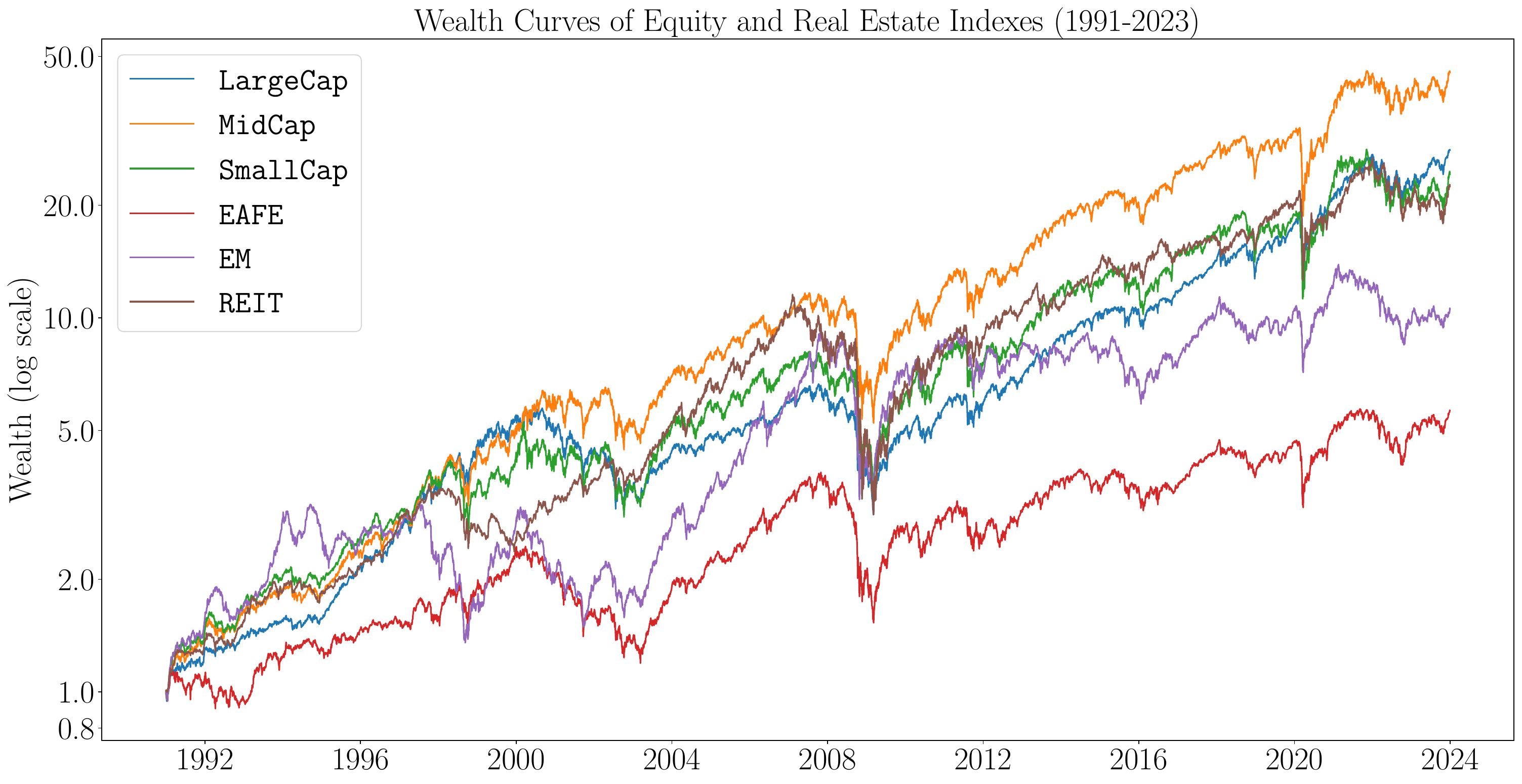}
    
    \includegraphics[width=\textwidth]{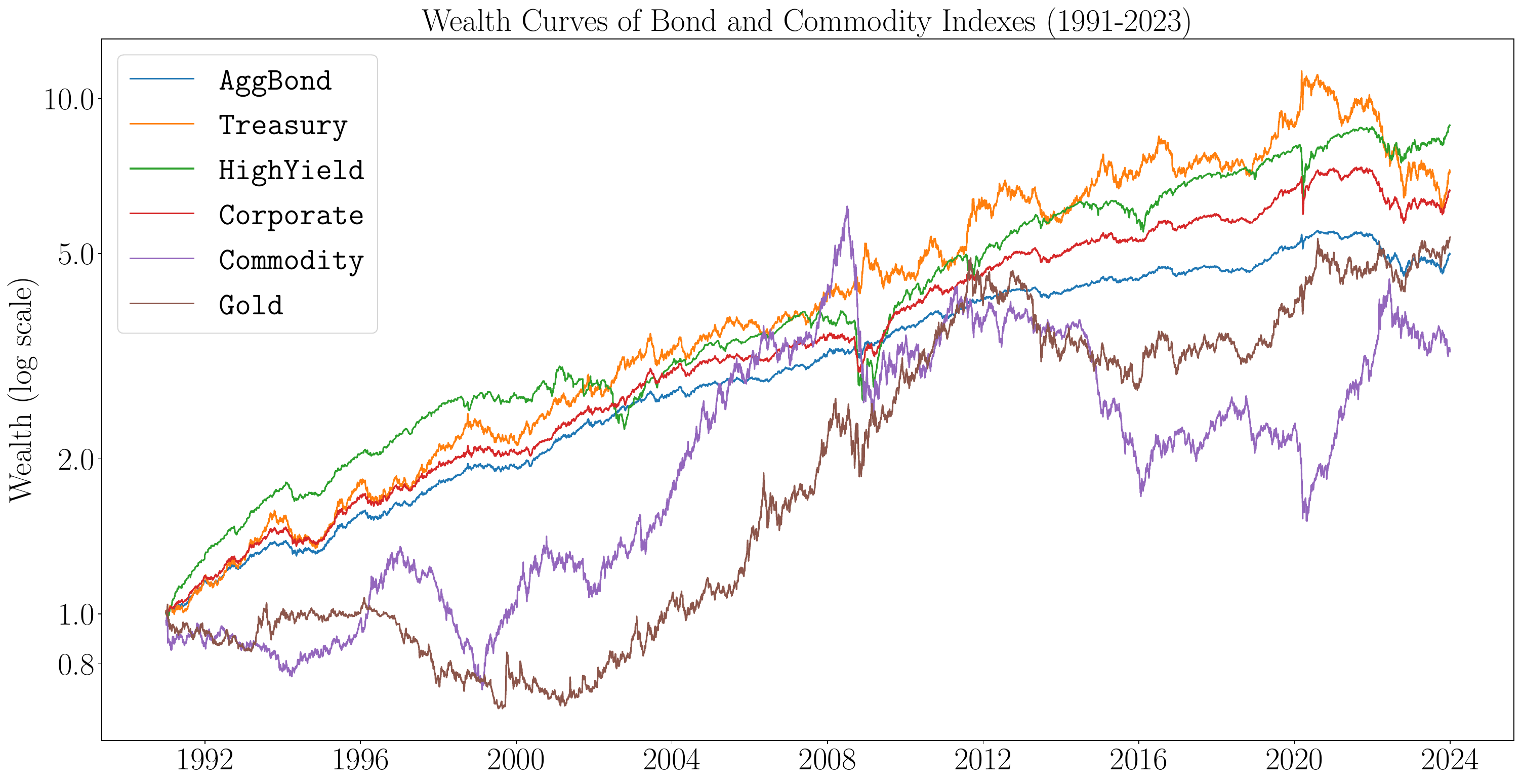}
    \caption{Wealth curves of the twelve asset classes from 1991 to 2023. 
    The upper figure displays equity and real estate indexes, and the lower figure displays bond and commodity indexes.
    Refer to \mbox{Table \ref{tab:assets}} for an overview of the asset universe.
}
    \label{fig:wealth all assets}
\end{figure}

Our asset universe offers extensive diversification across various countries and asset classes, including both developed and emerging markets, and spans all assets typically encountered by traditional allocators: equities, fixed income, real estate, and commodities. 
The diverse performance of these asset classes over the period from 1991 to 2023 is illustrated by the wealth curves of the 12 indexes, as shown in Figure \ref{fig:wealth all assets}.

In the equity and real estate indexes illustrated in the upper part of Figure \ref{fig:wealth all assets}, emerging markets show a relatively low correlation with developed markets, although this correlation has increased over time. 
Notably, \texttt{REIT} was not impacted by the dot-com bubble, which significantly affected other equity indexes.
Comparing the upper part with the lower part of Figure \ref{fig:wealth all assets}, which features bond and commodity indexes, the disparities are even more pronounced. 
Typically, during major market downturns such as the dot-com bubble, the global financial crisis, and the COVID crash, the well-documented ``flight-to-quality'' phenomenon \citep{Papadamou2021} causes stocks and bonds to move in opposite directions. 
However, this pattern did not occur during the 2022 market turbulence triggered by the Federal Reserve’s interest rate hikes, when both stocks and bonds suffered significant losses. 
The low or negative correlation of commodities with equities, along with their inflation-hedging characteristics, underscores their potential to offer substantial diversification benefits within a global portfolio \citep{Satyanarayan1996}.

Furthermore, the wealth curves reveal persistent bullish or bearish market regimes for nearly all assets, supporting the economic rationale behind our regime forecasts potentially enhancing portfolio construction. 
Despite the benefits of diversification, the varied behaviors of different assets present challenges to the robustness and generalizability of our proposed hybrid regime identification-forecasting framework, as it is applied individually to each asset.
Example plots of the regime forecasts generated by our framework for some assets, shown in Section \ref{subsec:regime examples}, demonstrate the method's ability to capture the distinct market regimes of various assets.

\section{Regime Identification and Forecasting}   \label{sec:regime}

Our methodology is structured as a multi-step process, primarily comprising regime identification, regime forecasting, and portfolio construction, with rigorous out-of-sample testing that accounts for transaction costs. 
The initial two steps are detailed in this section, while portfolio construction is discussed in the next section. 
For those who prefer a high-level overview first, you may skip to Section \ref{subsec:method summary} for a summarized description of the entire process.
A one-way transaction cost of 5 basis points is applied throughout the study.

Here, we start with an overview of the hybrid regime identification-forecasting framework applied to each individual asset, highlighting the benefits of employing separate models for each step. 
This is followed by detailed discussion of the identification and forecasting steps in their respective subsections, and our strategy for hyperparameter tuning. 
This section concludes with sample figures of the generated regime forecasts for selected assets to illustrate our methodology.

\subsection{Overview of the Regime Identification-Forecasting Framework}    \label{subsec:regime overview}

The primary outcome of the first two steps in our methodology is the generation of regime forecasts that can be realistically implemented in a live-sample setting. 
One significant challenge is the absence of directly observable regime labels for training, as regimes are predominantly unobservable and must be inferred for each historical period\footnote{
While there are some observed regime indicators, such as the business cycles dated by the National Bureau of Economic Research (NBER) used in \citet{james19}, their practicality for investment purposes is limited due to substantial lags in data availability and infrequent updates \citep{nystrup2015asset}. 
Furthermore, \cite{Uysal2021} demonstrates that using NBER-dated recessions as training targets underperforms compared to employing market regimes identified by a customized model for regime-based asset allocation.
}.   %
Therefore, every implementable regime-based asset allocation strategy  inherently divides into two distinct steps: identification and forecasting.  
Identification is an unsupervised learning problem where we assign a discrete label sequence $s_0,\ldots,s_{T-1}$ to the input (unlabeled) market data $\bmx_0,\ldots, \bmx_{T-1}$, describing the prevailing regime for each period. 
Here, features $\bmx_t$ are only known at the end of period $t$, making this an inherently interpretative (instead of predictive) process, as we seek to discern the pattern of period $t$ based on its  characteristics.
On the contrary, forecasting is a supervised learning task, where, given the market data $\tilde\bmx_{T-1}$\footnote{The tilde denotes that the features used in the two steps may differ.} (with its full history) up to the end of period $T-1$, we aim to generate a forecast $f_T$ for the upcoming trading period $T$, enabling timely asset allocation decisions. 
This clear separation provides more granular insights into each step, enhancing our understanding of regime-based asset allocation.

There are two major reasons why the distinction between regime identification and forecasting has not been extensively discussed in the literature.
First, the forecasting step is often simplified by directly using the label $s_{T-1}$, identified in the initial step using data available only up to period $T-1$, as the forecast $f_T$  for the next period.
This practice, commonly adopted in several studies including \citet{bulla2011asset, nystrup2015asset, Nystrup2018}, %
is generally acceptable due to the persistence inherently assumed in regime definitions, especially at a daily frequency where regime shifts are not expected to occur on a daily basis. 
Second, traditional Markov-switching models combine capabilities for both identifying historical patterns and forecasting future states by imposing likelihood assumptions on the data-generating process. 
Typically, a first-order homogeneous Markov chain governs the transitions between unobserved states, and state-conditional likelihoods define the distributional family for each state.

We adopt a formal approach to both regime identification and forecasting by employing two separate models for each, driven by three key reasons. 
First, using a distinct forecaster allows for a deeper understanding and better anticipation of regime shifts.
\citet{dacco1999} highlights the limitation of Markov-switching models’ predictive capabilities, in the context of forecasting exchange rates, finding that  ``[Markov-switching models] give good in-sample performance, but they are usually outperformed by [simpler baseline models] when used for forecasting'',  primarily due to misclassifications.
Second, the intrinsic differences between the unsupervised and supervised nature of the tasks make it challenging for a single model to excel in both. 
Unsupervised learning is notably more complex than supervised learning, due to the absence of class labels \citep{dash1997dimensionality} and the difficulties posed by high-dimensional feature spaces\footnote{
For example, the distance or similarity measure required by most clustering algorithms is heavily affected by the ``curse of dimensionality'', meaning that, as dimensionality increases, the distance to the nearest data point approaches the distance to the farthest point \citep{Beyer1999}, making it challenging to identify meaningful clusters. 
This issue is also relevant in high-dimensional (Gaussian) hidden Markov models, where the log-likelihood calculation involves squared Euclidean distances.
}
\citep{Steinbach2004}, among other reasons.
By forcing one model to perform both tasks, we sacrifice significant flexibility, including the ability to employ different feature sets tailored to each respective step.
Third, employing two separate models enables potential synergies, where the identification model can generate valuable labels that enhance the forecasting accuracy of the second model. 
Additionally, hyperparameter tuning across both steps can lead to better coordination and optimization, leveraging the strengths of each model.

Specifically, in the regime identification step, we derive regime labels using the statistical jump model, a temporal clustering algorithm utilizing features extracted from the asset return series.   
These labels are then shifted forward by one day and serve as prediction targets in the regime forecasting step, where a machine learning classifier utilizes a broader set of features.
This method is applied to each individual asset to generate asset-specific market regime forecasts.
This ``cluster-then-classify'' approach, which is effective in scenarios with scarce class labels and allows for a flexible choice of feature sets tailored to each task, has been successfully applied in various data science domains, including text classification \citep{Jiang2011}, biomedical informatics \citep{Peikari2018, GalvezGoicurla2022}, and predicting gas pipeline failures \citep{Alobaidi2022}.
In the realm of regime-based asset allocation, \citet{Uysal2021, Gu2021} explored a similar idea using an $\ell_1$ trend-filtering algorithm to identify historical regimes, coupled with a machine learning classifier for subsequent forecasting.
Our approach extends theirs by employing a more sophisticated regime identification model and conducting a more rigorous out-of-sample analysis\footnote{
In both studies,  regime labels were derived all at once by applying the trend-filtering algorithm to all data available through the end of the testing period. 
In contrast, we fit regime labels using only historical data available up to the current period and then use these labels to train our forecaster. 
This approach ensures that our strategy remains as realistic as possible.
}.

\subsection{Regime Identification}

It is essential to clarify the specific type of regime our study focuses on. 
The concept of a regime can be broad, inclusive, and somewhat elusive due to its unobservable nature and the versatility it encompasses. 
In regime-based asset allocation, previous studies have typically viewed regimes as a universal lens to understand  economic conditions that affect all asset performances, thereby assuming that a concise set of regimes can capture the time-varying behavior of all assets within a universe. 
For example, \citet{ang2004} developed a two-regime international CAPM model for six developed market equity indexes, characterized by a high-volatility bearish regime with spikes in international correlation. 
\citet{BAE2014} employed a multivariate hidden Markov model on return vectors of diverse assets including stock, bond and commodity indexes, aiming to generate regime-switching signals for asset allocation.
\citet{kim2023} analyzed two macro-indicators to establish four regimes, aiming to segment the historical performance across a similar range of assets. 
Furthermore, \citet{nystrup2015asset} extended the regimes identified from an equity index to a bond index, based on the premise that ``portfolio risk is typically dominated by stock market risk''.
While the adoption of a unified set of economic regime concepts across the entire asset universe offers benefits such as interpretability,
this method might not adequately account for the variation among asset classes, especially those with minimal correlation and driven by distinct fundamentals, such as equities and commodities.

In our regime identification approach, we prioritize two key requirements:
first, the identified regimes must be able to directly enhance the portfolio performance of asset allocation; 
second, these regimes should enable accurate forecasting in the subsequent step. 
These priorities guide our shift to identifying asset-specific market regimes using jump models, offering several benefits.
Firstly, we focus on market dynamics over broad economic conditions to more closely link identified regimes  with the  ultimate portfolio performance.  %
To emphasize this focus, we refer to  our identified regimes as \emph{market regimes} rather than \emph{economic regimes}. 
Secondly, by identifying regimes for each asset individually within the universe, we capture the nuances of every market, avoiding the simplification of using a major equity market like US \texttt{LargeCap} to represent all asset classes.  %
This targeted approach also mitigates the risk of incorrect regime forecasts by employing an ensemble of them.
Indeed, the consequences of incorrect regime forecasts can be severe; as \citet{hess2006} notes, ``a wrong regime forecast may not only lead to a non-optimal but to a detrimental allocation in the contrary direction''.
The third benefit arises from the fact that our regime forecasts, being asset-specific, are somewhat analogous to directly forecasting all asset returns. 
This similarity leads to a significant advantage: our identified regimes typically exhibit an enhanced signal-to-noise ratio (SNR) compared to direct return forecasting, thereby creating favorable conditions for the subsequent forecasting step.   %
This improvement stems from the persistence inherent in the regime definition, leading to regimes that effectively act as a heavily smoothed version of raw returns, for the purpose of forecasting\footnote{
Preprocessing raw returns in defining forecasting targets proves efficient for enhancing SNR. %
For instance, removing return components explained by factors, as noted by \citet{Kaniel2023}, leads to ``higher accuracy and better portfolio performance'' when predicting abnormal returns instead of total returns for mutual funds. 
Similarly, smoothing raw returns over a forward window, as implemented in \citet{aitsahalia2022}, results in a ``less noisy response variable'', which is ``more indicative of aggregated trading behavior [in a forward window] rather than at an arbitrary point in the near future.''
}.
Predictability varies across assets, and we fine-tune the jump penalty, which moderates the SNR, as outlined in Section \ref{subsec: hyperparam}, to optimize performance for each asset.

\subsubsection{Statistical Jump Models}  \label{subsec:JM}

In this article, we focus on the application of statistical jump models (JMs) as the regime identification model. 
While traditional Markov-switching models, notably hidden Markov models (HMMs), have been extensively developed and applied to financial return series since the seminal work of \citet{ryden1998},
 recent studies \citep{nystrup2020originalJump, nystrup2020onlineClass, Aydinhan2024} have highlighted their limitations.
For instance, the high imbalance and persistence of regimes, a low signal-to-noise ratio, and limited data availability often cause HMMs to generate state sequences that lack persistence and stability, leading to frequent false alarms.    %
Such inaccuracy could potentially undermine the effectiveness of regime-switching dynamics in asset allocation strategies \citep{shu2024regime}.
In response to these challenges, there has been a rise in research on non-parametric and data-driven alternatives, including the trend-filtering algorithm \citep{mulvey16} and the spectral clustering HMM (SC-HMM) \citep{zheng21}.%

\citet{Bemporad2018} introduced the jump model as a general unsupervised learning algorithm that fits multiple model parameters to time series data while incorporating temporal information. 
In our application, given an observation sequence of $D$ standardized features $\bmx_0,\ldots,\bmx_{T-1}\in \R^D$ derived from an asset return series $r_0,\ldots, r_{T-1}$ (feature engineering is detailed later), 
we estimate a \mbox{$K$-state}\footnote{
In general model discussion, terms \emph{state} or \emph{cluster} are used. 
Within our asset allocation application, they specifically denote a market regime.
} JM  by solving the optimization problem:
\begin{equation}\label{eq:jumpObj}
\min_{\Theta, \bmS}\quad\sum_{t = 0}^{T-1}l(\bmx_t, \bmtheta_{s_t})+\lambda\sum_{t = 1}^{T-1} \indicator{s_{t-1} \neq s_{t}}	\,. 
\end{equation} 
Here, the optimization variables are the $K$ model parameters $\Theta:= \lB{\bmtheta_k \in\R^D :k= 0, \ldots, K-1}$, and the unobserved state sequence $\bmS:=\{s_0,\ldots, s_{T-1} \}$. 
The optimal state sequence $\hat s_0,\ldots,\hat s_{T-1}$ serves as the prediction targets for the next forecasting step.
Each model parameter $\bmtheta_k$ acts as a representative for the \mbox{$k$-th} state, often referred to as cluster center/centroid in clustering analysis.
Each state variable $s_t$ takes one the $K$  discrete values in $\{0, 1, \ldots, K-1\}$, representing the assigned state for period $t$ and connecting features $\bmx_t$  with the corresponding model parameter $\bmtheta_{s_t}$, whose dissimilarity is quantified by a loss function $l(\cdot, \cdot)$.         %
Commonly, we use the scaled squared $\ell_2$-distance as the loss function, defined by  $l(\bmx, \bmtheta) := \frac12\lV[2]{\bmx- \bmtheta}^2$.

$\lambda\ge0$ is a critical hyperparameter known as the \emph{jump penalty}, moderating the fixed-cost regularization term, triggered whenever there is a jump between states. 
The objective function \eqref{eq:jumpObj} represents a balance between fitting the data with multiple model parameters and reflecting our prior belief about the persistence of the state sequence. 
With $\lambda = 0$, the model reduces to a \mbox{$k$-means} clustering algorithm, ignoring temporal dynamics. 
As $\lambda$ increases, state transitions become increasingly infrequent, eventually making even a single jump so costly that all data points are grouped into one cluster.
The method for selecting an optimal $\lambda$ is detailed later.

The optimization algorithm for \eqref{eq:jumpObj} employs a coordinate descent algorithm, as outlined in \citet{nystrup2020originalJump}, alternating between optimizing the model parameters $\Theta$ and the state sequence $\bmS$  while keeping the other variable fixed from the previous iteration\footnote{
An implementation of a collection of jump models, following the \texttt{scikit-learn} API style, is available on the first author’s GitHub page (\url{https://github.com/Yizhan-Oliver-Shu/jump-models}).
}. 
After fitting the JM, we calculate the transition probability matrix from the optimal hidden state sequence and compute metrics such as average return and volatility for each cluster.
We adopt a two-regime analysis with $K=2$ for its interpretability, distinguishing between bullish and bearish regimes based on the cumulative excess return. 
The regime with higher cumulative excess return is labeled as the bullish regime ($s_t=0$), while the one typically showing negative cumulative excess returns is considered bearish ($s_t=1$).
We update the jump model biannually, using an 11-year lookback window to recalibrate the model parameters $\Theta$ and the optimal state sequence $\bmS$.
The complete fitting scheme is discussed at the end of Section \ref{sec:forecasting}.

Beyond its basic function of controlling the persistence level, the jump penalty also plays a crucial role in moderating the signal-to-noise ratio (SNR), for our purpose of generating regime labels for subsequent forecasting. 
The optimal state sequence $\hat{s}_0,\ldots,\hat{s}_{T-1}$ acts as a smoothed version of the raw returns $r_0,\ldots,r_{T-1}$. 
With a low $\lambda$ value, the state sequence  jumps more frequently, promptly responding to changes in market dynamics, including those driven by  noise. %
Conversely, a high $\lambda$ value results in a smoother state sequence that is less sensitive to noise but experiences greater latency.
In the extreme case where $\lambda=0$ and raw returns are used directly as input features to JMs, \ietext $\bmx_t=r_t$, the model would classify a positive return $r_t$ as   bullish and a negative return as bearish\footnote{
The specific threshold of a two-cluster $k$-means algorithm applied to the return series might not be exactly zero. ``Positive/negative'' is for illustrative purposes only.
}. 
While these identified regimes may appear useful for asset asset allocation due to their immediacy, the inherent low SNR significantly challenges the subsequent forecaster's ability to accurately  predict the direction of future returns.
Choosing a non-zero $\lambda$ enhances the SNR of the identified labels, albeit at the cost of increased latency. 
The trade-off among accuracy, latency and SNR, as discussed in \citet{Nystrup2018}, is pivotal for the successful application of regime identification algorithms.
This article employs a time-series cross-validation method that evaluates strategy performance based on the combined outcomes from the identification-forecasting framework, as detailed in Section \ref{subsec: hyperparam}.

Our features used in JMs include eight financial measures calculated from an individual asset excess return series, designed to determine whether the asset is in a bullish or bearish market. 
Referred to as return features, these measures are asset-specific since a JM is fitted individually.   %
Specifically we utilize exponentially weighted moving (EWM)  downside deviation (DD) in log scale, average returns, and Sortino ratio across various halflives, as summarized in Table \ref{tab:jm feats}.
We prefer DD to volatility as a risk measure due to its emphasis on downside risk, and we apply a log transformation to DD measures to ensure stable model fitting, especially as volatility measures tend to spike during market turmoil. 
To alleviate feature correlation, we select only two halflives for DD since risk measures display much higher correlation across different halflives compared to return measures\footnote{
We exclude the two DD features for three assets: \texttt{AggBond}, \texttt{Treasury}, and \texttt{Gold}, as preliminary in-sample analysis indicates that risk features do not distinctly separate under the two fitted regimes. 
However, the forecasting algorithm described in Section \ref{sec:forecasting} includes all eight return features for all assets, thanks to its improved accommodation to high-dimensionality.
}. 
The EWM Sortino ratio, defined as the ratio of EWM average returns over \mbox{EWM DD}, represents risk-adjusted returns. 
For exponential smoothing, we choose halflives corresponding to one week, two weeks, and a month in trading days, avoiding shorter halflives due to increased noise.

    \begin{table}[htbp]
        \centering
        \begin{tabular}{c c}
        \toprule
             \textbf{Feature} & \textbf{Halflives (days)} \\
             \midrule
              Downside Deviation (log scale) & 5, 21 \\  
              \midrule
              Average Return & 5, 10, 21 \\ 
              \midrule
              Sortino Ratio & 5, 10, 21\\
              \bottomrule 
              \multicolumn{2}{l}{\rule{0pt}{2.75ex}   Note: Each feature is exponentially smoothed.}
        \end{tabular}
        \caption{List of eight return features used in jump models, exponentially smoothed over specified halflives, derived from an asset excess return series.}
        \label{tab:jm feats}
    \end{table}

Given the challenge of high-dimensional features pose to unsupervised algorithms, we select a succinct set of features derived solely from asset returns. 
The integration of cross-asset macro-features with these return features is deferred to the forecasting step.    %
Regarding the optimal model parameters, $\hat\bmtheta_0$ and $\hat\bmtheta_1$, the identified regimes are typically characterized by the bullish regime having lower risk  and higher return measures, while the bearish regime presents the opposite characteristics.

\citet{Bemporad2018} demonstrated that, under certain assumptions, JMs can nest HMMs within a probabilistic setting. 
Subsequently, \citet{nystrup2020originalJump} introduced JMs to financial applications, illustrating that JMs can outperform HMMs in statistical accuracy and can produce more persistent hidden state sequences.    %
Since then, several extensions to JMs have been developed. 
\citet{nystrup2021sparse} introduced a feature selection procedure to manage high-dimensional feature sets within the JM framework. 
\citet{Aydinhan2024} developed the \emph{continuous statistical jump model} (CJM), which generalizes the discrete hidden state variable into a probability vector across all states, offering a probabilistic interpretation where the hidden state vector indicates the probability of each period belonging to each regime. 
\citet{Cortese2024}  developed a generalized information criterion (GIC) to facilitate model selection in high-dimensional JMs, applying this methodology to cryptocurrency regime analysis with an extensive feature set \citep{cortese2023crypto}. 
\citet{Bosancic2024} proposed a heuristic for feature subset selection in JMs.
Additionally, \citet{Fantulin2024} introduced a regime identification model utilizing the recurrent neural network (RNN), inspired by the principles of JMs.

\subsection{Regime Forecasting}  \label{sec:forecasting}

With the optimal state sequence $\hat s_0,\ldots,\hat s_{T-1}$ identified by JMs on historical periods, the next step is to predict the upcoming regime  using all available information up to the current period. 
To achieve this, we train a classifier on the training dataset $\lB{\lp{\tilde \bmx_t, \hat s_{t+1}}}$, allowing us to make timely prediction of the prevailing regime $f_{T+1}$ for the next trading period when new market data $\tilde \bmx_T$ is available. 
Here, $\tilde{\bmx}_T$ represents an expanded feature set based on the return features $\bmx_T$ used by JMs, which will be detailed later. 
Our primary focus is on predicting the regime for the immediate next period. 
While there is flexibility to train forecasts for multiple prediction horizons, such as using $\tilde{\bmx}_t$ to predict the regime one week or one month later, \ietext $\hat{s}_{t+5}$ or $\hat{s}_{t+21}$, to support multi-period trading strategies \citep{Boyd2017}, it is important to note that extending the prediction horizon can substantially increase the forecasting difficulty.

For the classifier algorithm, we choose a gradient-boosted decision tree classifier implemented by XGBoost \citep{Chen2016}, recognized for its efficiency, scalability, and portability. 
Specifically, we train an \texttt{XGBClassifier} using the default hyperparameters to avoid the necessity for extensive tuning. 
Based on our experience, the algorithm proves robust and generally requires minimal hyperparameter tuning for our application. 
We use the probabilities predicted by the classifier and set a threshold of 0.5 to distinguish between forecasted bullish and bearish regimes.
Since the XGBoost classifier does not incorporate explicit temporal information, the persistence of the inferred regime probability series is reduced. 
To mitigate this and smooth out noise, we apply exponential smoothing to the probability series forecasted by XGBoost, selecting from a range of halflives\footnote{
We select a halflife from a list of candidates -- 0 (no smoothing), 2, 4, 8 days -- based on the performance of the \mbox{0/1} regime-switching strategy for each asset during the initial validation window from 2002 to 2007, as detailed in the next subsection.   %
Specifically, we choose a halflife of 8 days for: \texttt{LargeCap}, \texttt{MidCap}, \texttt{SmallCap}, \texttt{REIT}, \texttt{AggBond}, and \texttt{Treasury}; 4 days for: \texttt{Commodity} and \texttt{Gold}; 2 days for: \texttt{Corporate}; and 0 day for: \texttt{EM}, \texttt{EAFE}, and \texttt{HighYield}.
This selection is robust and exhibits a clustering phenomenon, where assets with similar fundamentals share the same halflife values.
}.

For the XGBoost classifier, we incorporate all eight return features $\bmx_t$ used in JMs, alongside a list of five cross-asset macro-features, with details provided in Table \ref{tab:macro-features}. 
While the return features are asset-specific, the same set of macro-features is applied to all assets. 
These macro-features are derived from four series sourced from FRED, all based on traded markets and available for real-time daily retrieval.
The first series is the market yield on US Treasury at 2-year constant maturity, serving as a proxy for interest rates, whose fluctuations significantly influence the performance of various assets, including stocks and bonds. 
For instance, the interest rate hikes by the Federal Reserve in 2022 led to notable losses across most equity and bond indexes. 
The second series is the spread between the US Treasury 10-year and 2-year yields, commonly known as the yield curve slope. 
This spread is widely recognized by both academics and financial media as a reliable predictor of future economic activity \citep{Estrella2006}.
The third series is the VIX index, calculated as the implied volatility of S\&P 500 index options and often referred to as the ``investor fear gauge'' due to its spikes during market turmoil \citep{Whaley2009vix}. 
The final series is the stock-bond correlation, calculated from the asset returns of \texttt{LargeCap} and \texttt{AggBond} in our dataset. 
This correlation is known to exhibit a time-varying pattern  across different macro-economic conditions \citep{Yang2009}.

\begin{table}[htbp]
\centering
\begin{tabular}{cc}%
\toprule
\textbf{Feature} & \textbf{Transformation} \\ 
\midrule
US Treasury 2-Year Yield & Difference and EWMA (hl=21) \\ \midrule
\makecell{Yield Curve Slope \\ (10-Year minus 2-Year)} & EWMA (hl=10) \\  \midrule
\makecell{Yield Curve Slope \\ (10-Year minus 2-Year)} & Difference and EWMA (hl=21) \\ \midrule
VIX Index & Log-difference and EWMA (hl=63) \\  \midrule
Stock-Bond Correlation & Rolling correlation (1-year lookback) \\ 
\bottomrule
\end{tabular}
\caption{List of five cross-asset macro-features used in the XGBoost classifier, combined with asset-specific return  features from Table \ref{tab:jm feats} for input into the classifier. 
All Treasury yields are based on constant maturity; 
``EWMA'' stands for exponentially weighted moving average, with ``hl'' specifying the halflife in trading days. 
Stock-Bond Correlation is calculated between the \mbox{S\&P 500} Index and the Bloomberg US Aggregate Bond Index.}
\label{tab:macro-features}
\end{table}

For feature transformation, we compute the exponentially weighted moving average (EWMA) difference of the 2-year Treasury yield to assess whether interest rates are in a rising or falling trend.
We also calculate the EWMA of the yield curve slope and its change to assess the current spread and whether it is widening. 
For the VIX index, we calculate the  EWMA of its log-differences to discern whether forward volatility is expected to increase.
Given the significant noise in daily VIX data, we use a higher halflife of 63 days to ensure smoother results.
While the absolute VIX index value, such as surges above 20.0, is frequently mentioned in financial media, we do not include it in our model due to its high correlation with our backward-looking downside deviation measures. 
Finally, we calculate the moving stock-bond correlation with a one-year lookback window.

Before delving into the optimal selection of the jump penalty in the next subsection, we here explain the fitting scheme of our hybrid framework to generate  asset-specific regime forecasts using a fixed jump penalty over a specific prediction window. Typically, this window spans several years, such as a validation period from 2002 to 2007.  %
For a specified jump penalty $\lambda$,  we biannually update the parameters for both the regime identification and forecasting algorithms, utilizing an 11-year lookback training window. 
Specifically, every six months beginning from the start of the prediction window, we first fit a 2-state jump model with the jump penalty $\lambda$, using the past 11 years of data, and then train an XGBoost classifier on this same 11-year window. 
We then use this classifier to perform daily online regime forecasts throughout the ensuing six-month period until the next model update.
These steps are implemented in parallel for all assets, allowing for the generation of realistic and implementable asset-specific regime forecasts for any given jump penalty in a live-sample environment.
Algorithm \ref{alg:regime} outlines our approach.
Given that our dataset begins in 1991, the initial training window spans from 1991 to 2002, allowing for the generation of forecasts under a fixed $\lambda$ value for any subsequent prediction period beginning after 2002.

\begin{algorithm}
\DontPrintSemicolon
\caption{Generation of asset-specific regime forecasts with our \texttt{JM-XGB} approach for a fixed jump penalty over a specified prediction window}  
\label{alg:regime}
\textbf{Input:}  Jump penalty $\lambda$, and a prediction window.

\For{each asset}{
    \For{every six months from the start of the prediction window}{
        \textbf{Fit the Jump Model:} Use features from Table \ref{tab:jm feats} over an 11-year training window.\;
        \textbf{Fit the XGBoost Classifier:} Use features from both Tables \ref{tab:jm feats} and \ref{tab:macro-features} over the same training window.\;
        \textbf{Execute Daily Forecasts:} Use the trained classifier to conduct daily online regime forecasts for the next six months.\;
    }
}
\textbf{Output:} Regime forecasts for all assets under the jump penalty $\lambda$ over the prediction window.
\end{algorithm}

\subsection{Hyperparameter Tuning}   \label{subsec: hyperparam}

We previously highlighted the importance of selecting the optimal jump penalty $\lambda$ for jump models in Section \ref{subsec:JM}.
Our methodology employs a time-series cross-validation approach, applied individually to each asset, with biannual updates of the optimal $\lambda$. 
Specifically, every six months from the start of the testing period, we use Algorithm \ref{alg:regime} to generate the regime forecasts over the preceding five-year validation window, for a range of candidate  $\lambda$ values.
These values range from 0.0 to 100.0, distributed evenly on a logarithmic scale.
Then we calculate the Sharpe ratio of a regime-switching strategy, referred to as the ``\mbox{0/1} strategy'', which leverages these regime forecasts over the validation period, for every considered $\lambda$.
This simple strategy, originally proposed by \citet{bulla2011asset}, alternates between 100\% investment in a single risky asset, or 100\% investment in the risk-free asset, depending on whether the forecasted regime for the risky asset is bullish or bearish, respectively, hence the name ``\mbox{0/1} strategy.''
We then select the $\lambda$ that yields the highest Sharpe ratio and maintain this value for the next six months to generate real out-of-sample regime forecasts post hyperparameter tuning.
This sequence of optimal jump penalties is also useful for computing the return forecasts needed for the Markowitz mean-variance optimization detailed in Section \ref{subsec:mv}. 
Given that our data begins in 2002, the first validation window spans from 2002 to 2007, with the out-of-sample testing period spanning from 2007 to 2023. 
Algorithm \ref{alg:jp select} outlines our approach.
We expand on the rationale of this approach from the following aspects.

\begin{algorithm}
\DontPrintSemicolon
\caption{Optimal jump penalty selection}
\label{alg:jp select}
\textbf{Input:} A list of candidate jump penalties.

\For{each asset}{
    \For{every six months from the start of the testing period}{
        \For{each $\lambda$}{
            \textbf{Generate Forecasts:} Utilize Algorithm \ref{alg:regime} to generate regime forecasts using the jump penalty $\lambda$ over a five-year lookback validation window.\;
            \textbf{Calculate Sharpe Ratio:} Compute the Sharpe ratio for the \mbox{0/1} strategy over the validation window.\;
        }
        \textbf{Select Optimal $\lambda$:} Identify the $\lambda$ that yields the highest Sharpe ratio and maintain this $\lambda$ for the ensuing six months.\;
    }
}
\textbf{Output:}  Sequences of optimal jump penalties and the corresponding out-of-sample regime forecasts over the testing period for all assets
\end{algorithm}

First and foremost, we prioritize financial metrics for our selection criterion over statistical metrics such as classification accuracy used by \citet{Uysal2021} or information criteria used by \citet{Cortese2024}.
Since regime labels are identified by our method, they may not accurately represent the ``ground truth'' necessary for calculating classification accuracy. 
On the contrary, the \mbox{0/1} strategy effectively evaluates the financial implication of the statistical accuracy of the regime forecasts for a specific asset.
A comprehensive discussion of the \mbox{0/1} strategy and its effectiveness in mitigating downside risk in asset management is detailed in \citet{shu2024regime}.
We use the Sharpe ratio of this strategy as our primary selection criterion, noting that typically the optimal $\lambda$ excels across various (risk-adjusted) return metrics, including cumulative return, Sharpe ratio, Sortino ratio, and information ratio. 
Throughout this article, the performance of the \mbox{0/1} strategy serves to evaluate financially the prediction accuracy of our regime forecasts, as presented in Section \ref{subsec:results individual}.   %

Secondly, we emphasize the importance of using a validation period distinct from the training period, unlike the methods in \citet{nystrup2020originalJump, Aydinhan2024} which maximize in-sample accuracy, or \citet{Bosancic2024}, which optimizes in-sample cumulative returns. 
By utilizing a separate validation period, we align our hyperparameter tuning process closely with what would occur in a live-sample setting, thus enhancing the robustness of our methodology and preventing any potential look-ahead bias.

Other benefits include that tuning the optimal $\lambda$ separately for each asset achieves varied signal-to-noise ratios tailored to their distinct predictability levels.
This approach also avoids the complexity of tuning all twelve $\lambda$ values simultaneously, which potentially involves  a vast search grid. 
Additionally, $\lambda$ is tuned based on the regime forecasts, optimizing the synergy and coordination within the joint identification-forecasting framework to maximize their combined effectiveness.

We refer to our method of generating regime forecasts under the hybrid framework, following optimal hyperparameter tuning, as ``\texttt{JM-XGB}'' in the upcoming performance tables and plots in Section \ref{sec:results},  reflecting our joint approach that integrates the two distinct algorithms. 
As discussed in Section \ref{subsec:regime overview}, another viable approach is to directly use the identified regime $\hat s_{T-1}$ from the JM as the forecast $f_T$ for the next period, referred to simply as ``\texttt{JM}''.
We provide a comparison of the regime forecasts generated by the simpler \texttt{JM} approach against our comprehensive \texttt{JM-XGB} approach for all assets in Section \ref{subsec:results individual}.

\subsection{Illustrative Examples of Regime Forecasts}   \label{subsec:regime examples}

\begin{figure}[htbp]
    \centering     %
    \includegraphics[height=.29\textheight]{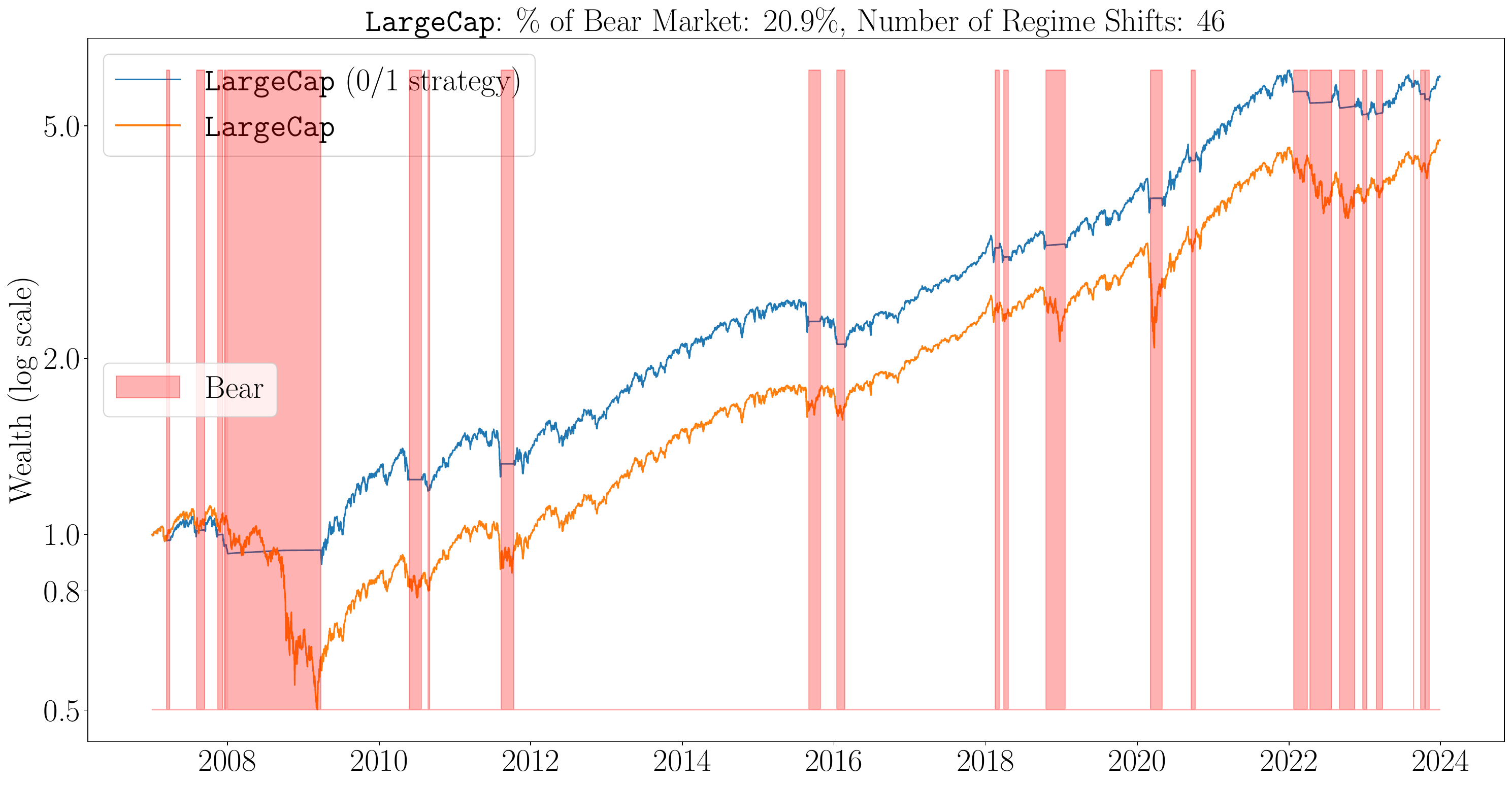}

    \includegraphics[height=.29\textheight]{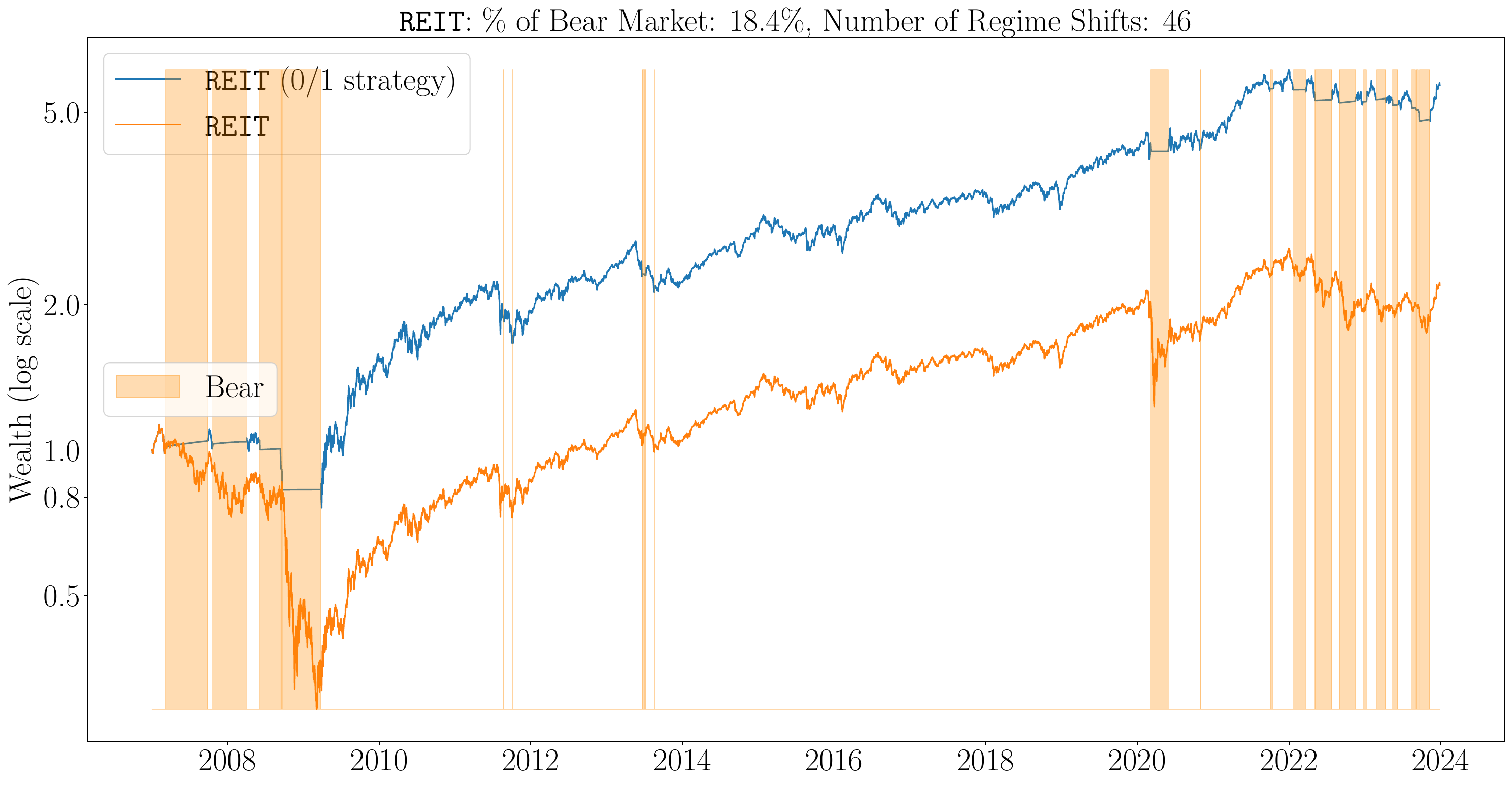}

    \includegraphics[height=.29\textheight]{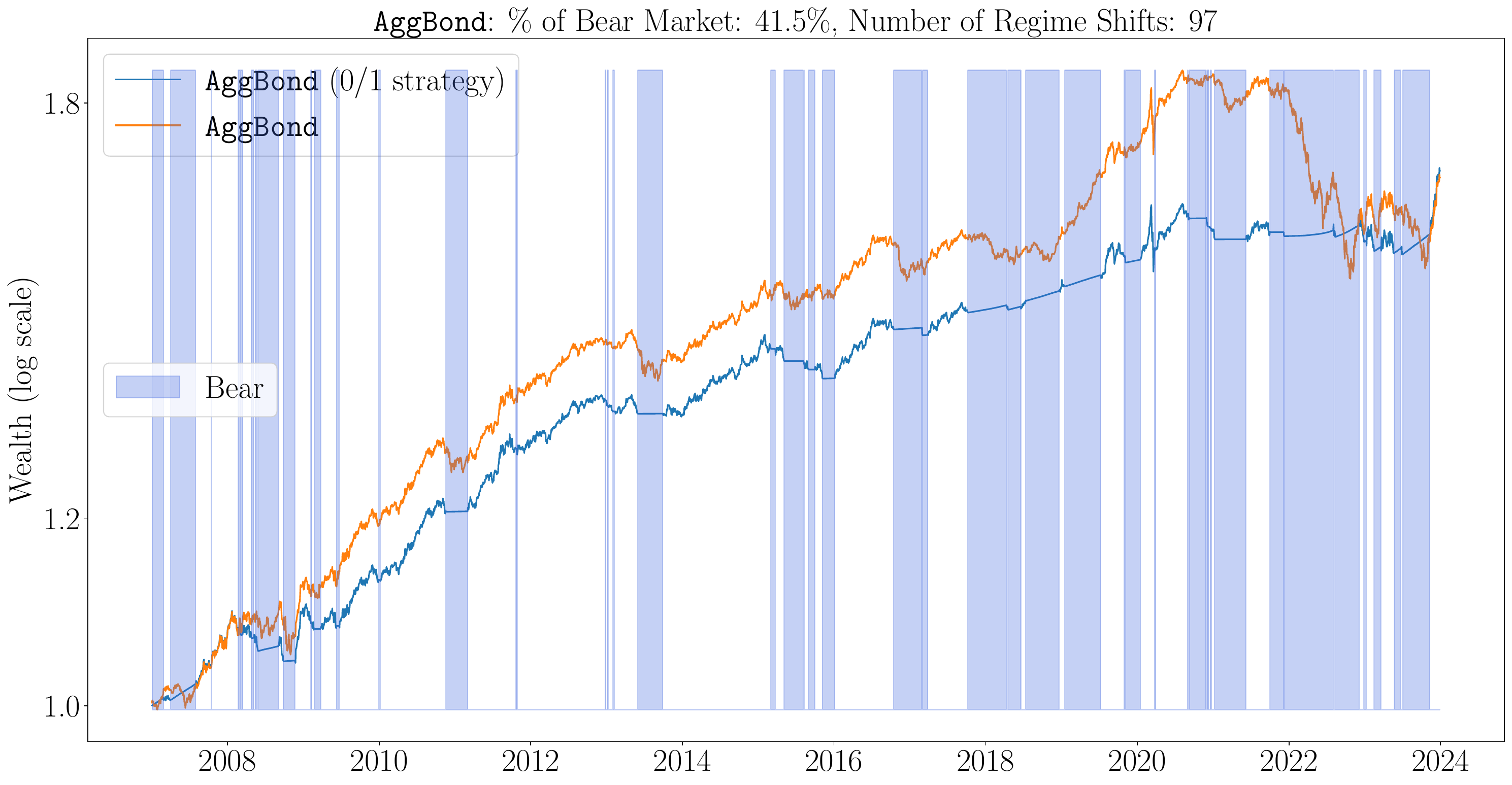}
    
    \caption{Regime Forecasts by our hybrid \texttt{JM-XGB} framework over the testing period (2007-2023) for three assets: \texttt{LargeCap}, \texttt{REIT}, and \texttt{AggBond}, displayed from top to bottom. 
    Colored shaded areas indicate forecasted bearish markets, during which the 0/1 strategy allocates 100\% to the risk-free asset.
    Otherwise, the strategy maintains 100\% investment in the risky asset.
    Yellow and blue curves represent the wealth curves of the asset and the \mbox{0/1} strategy applied using our regime forecasts, respectively. 
    A one-way transaction cost of 5 basis points is applied.
    }
    \label{fig:regime examples}
\end{figure}

To illustrate and visualize the outcomes of our hybrid regime identification-forecasting framework, we present figures of regime forecasts for three diverse assets: \texttt{LargeCap}, \texttt{REIT}, and \texttt{AggBond}, over the testing period from 2007 to 2023, as shown in Figure \ref{fig:regime examples}. 
The colored shaded areas indicate forecasted bearish market regimes, during which the 0/1 strategy would switch to 100\% in the risk-free asset. 
Yellow and blue curves represent the respective wealth curves of the asset and the \mbox{0/1} strategy applied using our regime forecasts.

The forecasted bearish regimes  notably capture major market downturns, such as the global financial crisis and the COVID crash for \texttt{LargeCap} and \texttt{REIT}, as well as the 2022 interest rate hikes affecting \texttt{AggBond}, during which the 0/1 strategy avoids substantial losses.
Additionally, our forecasts capture the high-volatility periods for \texttt{LargeCap} and \texttt{REIT} during the 2022 rate hikes. 
Consequently, the wealth curves for the 0/1 strategy finish either higher or comparable to those of a buy-and-hold strategy but with reduced volatility and drawdowns, as detailed in Section \ref{subsec:results individual}.
This emphasizes the 0/1 strategy's objective of downside protection and volatility mitigation while enhancing risk-adjusted returns.

Examining the regimes of \texttt{LargeCap}, \texttt{REIT}, and \texttt{AggBond} together highlights their distinctive nature and specific response to market dynamics. 
For instance, the bearish market of \texttt{REIT} preceded that of \texttt{LargeCap} in the 2008 financial crisis due to the earlier impact of the subprime mortgage crisis on real estate. 
The opposite regimes forecasted for the two indexes with \texttt{AggBond} during 2008 and 2020 provided diversification when investors  need it most. 
These demonstrates the advantage of considering asset-specific regimes.
The bearish forecasts for \texttt{AggBond} constitute 42\% of the periods, considerably higher than the 21\% for \texttt{LargeCap} and 18\% for \texttt{REIT}. 
This discrepancy primarily stems from our methodology of creating features based on each asset's excess return. 
During periods of high interest rates, the risk-free rate can be comparable to the return of \texttt{AggBond}. 
Notably, during the bearish periods forecasted for \texttt{AggBond} in 2018--2019 and 2023, the wealth curve of the 0/1 strategy shows a discernible upward trend, even though it was fully invested in the risk-free asset, reflecting the high interest rates at those times.

Admittedly, the regime forecasts are not perfect.
A prominent challenge is the delay of the forecasts, as, for example, observed in the 2015--2016 market selloff for \texttt{LargeCap}, influenced by complex global events including Chinese stock market turbulence, the Greek debt default, and Brexit.
These events highlight the challenge in providing timely bearish forecasts and underscore the necessity for incorporating more comprehensive data to improve forecast accuracy.

\section{Portfolio Construction} \label{sec:portfolio}

The primary objective of portfolio construction is to transform forecasting signals into optimized asset allocation weights. 
This section demonstrates the process for three distinct portfolio models: minimum-variance (MinVar), mean-variance (MV), and equally-weighted (EW) portfolios, each of which seamlessly incorporates our asset-specific regime forecasts. 
Before we explore each model, we first present the mathematical formulation of our Markowitz mean-variance optimization (MVO). 
Finally, Section \ref{subsec:method summary} provides a comprehensive summary of our methodology, as discussed in Section \ref{sec:regime} and \ref{sec:portfolio}.
We use the suffix ``(\texttt{JM-XGB})'' to refer to the models that incorporate our regime forecasts.

\subsection{Optimization Formulation}

We utilize a basic form of MVO as initially proposed by Markowitz, enhancing it with additional objective terms and constraints\footnote{
In the  1952 Markowitz article, the optimization problem  is equivalent to maximizing the risk-adjusted return $\bmw\tran \bmmu -\gamma^{\text{risk}} \bmw\tran\bmSigma \bmw$, under long-only and fully-invested weight constraints. Our modifications are: the addition of a trading cost term, the addition of an upper bound on weights, and replacing the leverage equality constraint with an inequality constraint.
}.
The optimization problem is formulated as follows:
\begin{align}
    \underset{\bmw}{\text{maximize  }}   \quad & \bmw\tran \bmmu -\gamma^{\text{risk}} \bmw\tran\bmSigma \bmw -\gamma^{\text{trade}} \times a \lV[1]{\bmw-\bmw^{\text{pre}}}     \label{eq:optimizer obj} \\
    \text{subject to}\quad                     & 0\le\bmw\le w^{\text{ub}},   \label{eq:optimizer weight constr}\\
                                               & \bfone\tran\bmw \le L.      \label{eq:optimizer lev constr}
\end{align}
Here, the optimization variables $\bmw$ represent the portfolio weights on all risky assets. 
$\bmmu$ and $\bmSigma$ are our forecasts for excess returns and their covariance matrix, respectively. 
$\gamma^{\text{risk}}$ denotes the risk aversion parameter.   
The term $a$ stands for the one-way transaction cost, set at 5 basis points.    %
$\bmw^{\text{pre}}$ denotes the pre-trade weights at the end of the previous period, with $\lV[1]{\bmw-\bmw^{\text{pre}}}$ quantifying the required trading  for rebalancing. 
Consequently, $a \lV[1]{\bmw-\bmw^{\text{pre}}}$ calculates the trading cost under a linear transaction cost, assumed for simplicity.
The trade aversion parameter, $\gamma^{\text{trade}}$, reflects our preference for limiting trading within a period. 
We enforce a long-only constraint and an upper bound $w^{\text{lb}}=40\%$ on the weight vector, along with a leverage limit of $L=1$, implying no short position in the risk-free asset. 
Since $\bmw$ only covers risky assets, a weight of $1-\bfone\tran\bmw$ is allocated to the risk-free asset when $\bfone\tran\bmw < 1$. 
At the end of each trading day in the testing period, we generate out-of-sample regime forecasts for all assets for the upcoming period and integrate these forecasts into the MVO to determine the desired allocation weights in various ways.
The methodology, including the choice of the trading parameters $\gamma^{\text{risk}}$ and $\gamma^{\text{trade}}$, is outlined in the following subsections.

We employ an enhanced version of MVO due to its originality and the ease of incorporating asset-specific regime forecasts, as shown later.   
If forecasts are only available for a subset of assets or if a reference portfolio plays a significant role, the Black-Litterman model, popular among practitioners, provides a viable alternative \citep{black1992}. 
Its use as a portfolio construction method in dynamic asset allocation approaches have been demonstrated by \citet{kim2023, Elkamhi2023}.

A significant amount of enhancements to the original  MVO formulation have been proposed. 
For a comprehensive review, we refer to the recent work by \citet{boyd2024markowitz}. 
Notably, it has become clear that incorporating appropriate objective terms and constraints is crucial for enhancing the practical application of MVO. 
These adjustments not only reflect the practical constraints and considerations of a portfolio manager, but also serve a similar purpose to the use of regularization techniques in machine learning.
While including these terms might worsen in-sample performance, they hold the potential to improve out-of-sample performance \citep{boyd2024markowitz}.  
This advantage is particularly relevant given the generally low signal-to-noise ratio in financial forecasts.
Modern developments in computational convex optimization have facilitated the incorporation of a diverse range of objective terms and constraints, such as forecasting error terms, turnover limits, and factor structures, within reasonable computing times.

In our implementation, we use a simple approach to handle transaction costs in single-period portfolio optimization by incorporating a trading cost objective term, as suggested by \citet{Boyd2017}, which prevents excessive trading and potentially reduces the model's sensitivity to noise in the return forecasts.    %
While there are arguably better alternatives for dealing with trading costs by  multi-period models \citep{li2022}, our addition provides a minimal yet effective solution. 
Regarding weight constraints, numerous studies, including  \citet{Frost1988, Jagannathan2003, Levy2014}, have demonstrated their effectiveness in enhancing out-of-sample performance for both MV and MinVar portfolios by acting similar to shrinkage estimation on the covariance matrix.

The MVO formulation we use leads to a Quadratic Programming (QP) problem.
It is Harry Markowitz who first introduced mathematical optimization to portfolio selection, effectively balancing the mean-variance tradeoff.
In the 1952 article, he solved the complete efficient frontier through geometric interpretation, due to the lack of established optimization algorithms.
Spurred by this innovation, there has been significant progress in the development of fast and efficient QP algorithms, facilitated by both open-source and commercial software. 
This evolution traces back to the early work of the Critical Line Algorithm by Markowitz \citep{markowitz56} and the Frank-Wolfe algorithm \citep{frank1956}, extending to the interior-point methods \citep{karmarkar84, Nesterov94} that are predominantly used by modern optimization software. 
In our implementation, we utilize the \texttt{Gurobi} optimizer via its Python modeling interface, \texttt{gurobipy}. 
Our tests show that running one pass of daily frequency backtesting, which involves solving thousands of sequential QPs for twelve assets, takes less than 5 seconds on a standard personal laptop, displaying satisfactory efficiency.

\subsection{Minimum-Variance Portfolio}   \label{subsec:minvar}

MinVar portfolio, identified as the leftmost point on the efficient frontier given risk on the $x$-axis, offers the practical advantage of not requiring return forecasts, which are notoriously difficult to estimate accurately. 
Instead, it relies on the more reasonably estimated covariance matrix, often refined through factor models and shrinkage methods.
More recently, MinVar strategy has been recognized as a market anomaly. 
\citet{ang2006} observed that stocks with higher historical idiosyncratic volatility tend to yield lower realized returns, and \citet{Clarke2006} noted that MinVar portfolio effectively reduces risk without compromising returns, leading to an improved risk-adjusted return compared to the market. 
Nowadays, the growth of exchange-traded funds (ETFs) has democratized access to sophisticated optimization-based investment strategies for retail investors. 
For instance, the ETF \texttt{USMV}, which employs a minimum volatility strategy across a broad range of US equities, had a market value exceeding 24 billion dollars at the time of writing.

Our implementation of the original MinVar portfolio adheres to the MVO formulation in \eqref{eq:optimizer obj}, by inputting constant return forecasts, \egtext $\bmmu\propto\bfone$, into the optimizer.
This is equivalent to the traditional approach of directly minimizing portfolio variance subject to constraints. 
For example, if $\bmmu=\bfone$ is inputted, and the trading cost term in \eqref{eq:optimizer obj} and weight constraints \eqref{eq:optimizer weight constr} are ignored, the optimal solution  becomes the standard MinVar portfolio $\bmSigma^{-1}\bfone$, scaled to a leverage of $L$, given $\gamma^{\text{risk}}$ smaller than a threshold of $\bfone\tran\bmSigma^{-1}\bfone/2L$.
If $\gamma^{\text{risk}}$ exceeds this threshold, the solution remains proportional to $\bmSigma^{-1}\bfone$ but with reduced leverage.

This approach facilitates the seamless integration of our regime forecasts into the optimization process.
Unlike previous regime-switching risk-based portfolio strategies, such as that proposed by \citet{Uysal2021}, which modifies the covariance matrix estimation based on regime information, our methodology involves adjusting the return forecast $\bmmu$. 
Specifically, we assign $\mu_j=10$ basis points to all assets forecasted to be in the bullish market, and set $\mu_j=0$ for assets forecasted as bearish.
This is equivalent to solving MinVar optimization within all assets predicted as bullish.
Since our MVO is long-only, asset with non-positive return forecasts generally receive no allocation.
Additionally, if fewer than four (three or fewer) assets are forecasted as bullish, we allocate 100\% to the risk-free asset to avoid risk concentration in a limited number of assets.

For covariance estimation, we use an exponentially weighted moving historical covariance matrix with a 252-day halflife.
We incorporate regime forecasts into return rather than risk forecasts, mainly due to the considerably larger impact that return forecasts have on MVO \citep{Chopra1993}.
This method underscores the benefits of regime-based asset allocation conservatively, while leaving the thorough integration of asset-specific regime forecasts into covariance estimation for future research.

Regarding trading parameters, as MinVar optimization is not sensitive to  the risk aversion parameter, we use $\gamma^{\text{risk}}=10.0$ across both MinVar and MV portfolios for consistency.
We exclude the trading cost term from the original MinVar portfolio due to its inherently low turnover, and apply  $\gamma^{\text{trade}}=1.0$ for the MinVar (\texttt{JM-XGB}) portfolio that incorporates our regime forecasts.

\subsection{Mean-Variance Portfolio}   \label{subsec:mv}

Despite the well-known challenges of applying  MVO in practice, its widespread use across the investment industry is undeniable. 
MVO remains to be a cornerstone tool for sophisticated investors like active portfolio managers, helping translate various alpha signals into desired portfolio positions \citep{grinold2019}, or serving as a basis for making asset allocation decisions \citep{Kim2021}.
Recent studies, including  \citet{boyd2024markowitz} and \citet{Benveniste2024}, have addressed and provided solutions for many of the perceived shortcomings of MVO.

Accurate return forecasts are crucial for successful MVO implementation. 
In our baseline MV portfolio, we employ admittedly simplistic return estimates based on an exponentially weighted moving average with a five-year halflife.
It's important to note that Harry Markowitz never suggested such naive estimates as adequate inputs for MVO; they are used here merely for comparative purposes. 
For our enhanced MV (\texttt{JM-XGB}) portfolio that incorporates regime forecasts, we derive our return forecasts from a combination of regime forecasts, regime-dependent historical returns calculated by the JMs, and the optimally selected jump penalty.
During in-sample training, each JM fitting calculates the average returns for periods identified as bullish or bearish. 
Then, when making online forecasts, we use the average return from historical periods that were in the same regime as forecasted for the next day, using the JM with the optimally selected jump penalty. 
For instance, if tomorrow is forecasted to be bullish with the optimally selected jump penalty $\hat{\lambda}$, we identify all past periods in the last training window classified as bullish from the JM under this penalty $\hat{\lambda}$, and use their average return as the forecast for tomorrow.
To prevent extreme values, we cap the return forecasts under a predicted bearish regime at negative 10 basis joints. 
A detailed statistical assessment of the accuracy of our regime-aware return forecasts is provided in Section \ref{subsec:portfolio res}.

Apart from the return forecasts, most other components of the optimizer are consistent with the MinVar portfolio.
For the MV (\texttt{JM-XGB}) portfolio, we set $\gamma^{\text{risk}}=10.0$, and for the original MV portfolio, $\gamma^{\text{risk}}=5.0$. 
We use a higher risk aversion for the regime-aware portfolio, mainly because forecasts under a specific regime generally display higher absolute values compared to regime-agnostic return forecasts; the latter can be thought of as a weighted average of forecasts under all regimes.
We set $\gamma^{\text{trade}}=1.0$ for the MV (\texttt{JM-XGB}) portfolio, while for the original MV portfolio, $\gamma^{\text{trade}}=0.0$, due to its already low turnover. 
A sensitivity analysis on both trading parameters is provided in Section \ref{subsec:sense}. 
We use an EWM historical covariance matrix with a 252-day halflife for our covariance forecasts. 
Additionally, if three or fewer assets are predicted to be in the bullish regime, we allocate 100\% to the risk-free asset to avoid concentrated exposure to a small number of assets.   %

\subsection{Equally-Weighted Portfolio}

EW portfolio, which distributes capital equally across a universe of assets, serves as a notable example of a naive diversification rule that does not depend on any forecasting or optimization skill. 
Despite debates sparked by a series of articles including \citet{demiguel2009optimal, Kirby2012, behr2103}, there is a consensus that, if not properly implemented, MVO-based portfolio methods including MinVar and MV, can easily underperform simple diversification approaches like EW in an out-of-sample setting, despite their theoretical soundness. 
Beyond its frequent use as a benchmark in academic research, the practical implementation of EW strategies has a rich history, ranging from the first index fund established in 1971 based on an equally-weighted index of New York Stock Exchange equities, to the popularity of ETFs such as \texttt{RSP}, which applies equal weights to the stocks in the S\&P 500 Index\footnote{
Based on a speech by John C. Bogle about the origin of indexing, retrieved from \url{https://boglecenter.net/wp-content/uploads/Superbowl-12-5-01.pdf}. 
The 1971 fund was designed for a pension, while the fund Bogle created is recognized as the first index mutual fund. 
\texttt{RSP} had a market value of over 55 billion dollars at the time of writing.
}.

In our implementation, the original EW portfolio allocates a fixed fraction of 1/12 to each of the twelve risky assets, with daily rebalancing. 
While daily rebalancing poses practical challenges, for the purpose of this benchmark analysis, the frequency of rebalancing does not significantly impact performance. 
For the EW (\texttt{JM-XGB}) portfolio, we distribute the entire 100\% weight equally among all assets forecasted to be   bullish and rebalance daily. 
Additionally, if fewer than four assets are predicted to be bullish, we invest 100\% in the risk-free asset to avoid risk concentration.

\subsection{Summary of Methodological Steps}    \label{subsec:method summary}

Here, we summarize our entire multi-step methodology as described in Section \ref{sec:regime} and \ref{sec:portfolio}, for the convenience of our readers. 
This summary references specific tables, algorithms and equations from the previous sections for more in-depth descriptions.

Algorithm \ref{alg:regime} outlines the fitting scheme of our hybrid regime identification-forecasting framework to generate asset-specific regime forecasts using a fixed jump penalty over a specific prediction window. 
We biannually update the parameters for both the regime identification and forecasting algorithms, utilizing an 11-year lookback training window. 
Specifically, every six months beginning from the start of the prediction window, we first fit a 2-state jump model with the specified jump penalty, using the feature set from Table \ref{tab:jm feats} over the past 11 years of data, to classify training periods into bullish or bearish states.
The derived regime labels are then shifted forward by one day to serve as prediction targets for the XGBoost classifier. 
We next train this classifier using features from both Table \ref{tab:jm feats} and \ref{tab:macro-features} over the same 11-year period. 
This classifier is then utilized to perform daily online regime forecasts over the ensuing six months until the next model update.
The initial training window spans from 1991 to 2002.
These steps are implemented in parallel for each asset.

Following this, Algorithm \ref{alg:jp select} outlines the optimal selection of the jump penalty through a time-series cross-validation method, applied individually to each asset with biannual updates.
Every six months starting from the testing period, we first use Algorithm \ref{alg:regime} to generate regime forecasts over the preceding five-year validation window for a series of $\lambda$ values, ranging from 0.0 to 100.0, distributed evenly on a logarithmic scale.  
We then evaluate the Sharpe ratio of the \mbox{0/1} strategy for each $\lambda$, choosing the jump penalty that results in the highest Sharpe ratio and maintaining this value for the next six months. 
This method produces sequences of optimal jump penalties and corresponding out-of-sample regime forecasts post hyperparameter tuning for all assets throughout the testing period from 2007 to 2023, with the first validation period from 2002 to 2007.

The regime forecasts are then integrated into various portfolio construction methods. 
For MinVar and MV portfolios, we use the MVO formulation in \eqref{eq:optimizer obj}. 
Covariance estimate $\bmSigma$ is chosen as the EWM historical covariance matrix with a 252-day halflife.
For MinVar portfolio, we uniformly input a return forecast of $\mu_j=10$ basis points for all assets forecasted to be bullish, and $\mu_j=0$ for those forecasted as bearish.
For MV portfolio, we use as the return forecast the average return from historical periods in the last training window that were in the same regime as forecasted for the next day, using the JM with the optimally selected jump penalty.
For EW portfolio, we allocate the  100\% weight equally among all assets forecasted to be bullish and rebalance daily.
For all models, we shift all investment to the risk-free asset if three or fewer assets are forecasted as bullish.

\section{Empirical Results}        \label{sec:results}

This section presents the empirical results of strategy performance over the testing period from 2007 to 2023. 
In Section \ref{subsec:results individual}, we detail the performance of the \mbox{0/1} regime-switching strategy applied to each individual asset, aimed at assessing the financial implication of the accuracy of our asset-specific regime forecasts. 
Section \ref{subsec:portfolio res} then presents the performance of dynamic asset allocation strategies that incorporate our regime forecasts through three distinct portfolio models: minimum-variance (MinVar), mean-variance (MV), and equally-weighted (EW) portfolios. 
A sensitivity analysis follows to evaluate the robustness of these results.
Throughout the empirical study, a one-way transaction cost of 5 basis points is applied.

It is worth emphasizing that our primary objective is to demonstrate how the integration of asset-specific regime forecasts generated by our methodology can enhance portfolio construction, rather than performing a comparative study of the three portfolio models, which has been extensively studied in literature such as \citet{demiguel2009optimal, bessler2017}.
We choose these models -- each offering distinct risk-return tradeoff to suit various investor preferences -- to translate our forecasting signals into asset allocation weights. 
Our analysis aims not to declare any portfolio model superior but to illustrate how each can benefit from our tailored regime forecasts.

\subsection{Regime-Switching Strategy Performance for Individual Assets}     \label{subsec:results individual}

This subsection examines the quality of our asset-specific regime forecasts, critical for transforming forecasts into meaningful portfolio weights. 
Previously in Section \ref{subsec: hyperparam}, we emphasized our preference for financial metrics over statistical classification accuracy by examining the \mbox{0/1} strategy, which alternates between  100\% investment in either a risky asset or a risk-free asset, depending on whether the predicted regime is bullish or bearish.
Table \ref{tab:perf indiv assets}  displays the Sharpe ratio and the maximum drawdown for this strategy applied to each asset during the 2007--2023 testing period, presented in the upper and lower sections of the table, respectively. 
The first row ``\texttt{B\,\&\,H}'' represents the buy-and-hold strategy, which serves as the \mbox{0/1} strategy under a dummy all-time bullish forecaster;
the second row ``\texttt{JM}'' represents the \mbox{0/1} strategy using regime forecasts derived solely from the jump model by carrying forward today's identified regime as tomorrow's forecast; 
the third row ``\texttt{JM-XGB}'' represents the \mbox{0/1} strategy using our hybrid framework, integrating jump models for initial regime identification and XGBoost for subsequent forecasting.

\begin{table}[htbp]
    \centering
\begin{tabular}{l*{6}c}
\toprule
\textbf{Sharpe Ratio}  & \texttt{LargeCap} & \texttt{MidCap} & \texttt{SmallCap} & \texttt{EAFE} & \texttt{EM} & \texttt{REIT} \\
\midrule
\texttt{B\,\&\,H}  & 0.50 & 0.45 & 0.36 & 0.20 & 0.20 & 0.27 \\
\texttt{JM} & 0.59 & 0.49 & 0.28 & 0.28 & 0.65 & 0.39 \\
\texttt{JM-XGB} & 0.79 & 0.59 & 0.51 & 0.56 & 0.85 & 0.56 \\
\midrule
 & \texttt{AggBond} & \texttt{Treasury} & \texttt{HighYield} & \texttt{Corporate} & \texttt{Commodity} & \texttt{Gold} \\
\midrule
\texttt{B\,\&\,H}  & 0.46 & 0.26 & 0.67 & 0.54 & 0.03 & 0.43 \\
\texttt{JM} & 0.43 & 0.21 & 1.49 & 0.83 & 0.08 & 0.12 \\
\texttt{JM-XGB} & 0.67 & 0.38 & 1.88 & 0.76 & 0.23 & 0.31 \\
\bottomrule
\multicolumn{7}{c}{ } \\ %
\toprule

\textbf{Max Drawdown}  & \texttt{LargeCap} & \texttt{MidCap} & \texttt{SmallCap} & \texttt{EAFE} & \texttt{EM} & \texttt{REIT} \\
\midrule
\texttt{B\,\&\,H} & -55.25\% & -55.15\% & -58.89\% & -60.41\% & -65.25\% & -74.23\% \\
\texttt{JM} & -24.78\% & -33.24\% & -38.35\% & -29.72\% & -26.22\% & -54.71\% \\
\texttt{JM-XGB} & -17.69\% & -29.89\% & -35.84\% & -19.93\% & -21.30\% & -32.70\% \\
\midrule
 & \texttt{AggBond} & \texttt{Treasury} & \texttt{HighYield} & \texttt{Corporate} & \texttt{Commodity} & \texttt{Gold} \\
\midrule
\texttt{B\,\&\,H} & -18.41\% & -46.91\% & -32.87\% & -22.04\% & -75.54\% & -44.62\% \\
\texttt{JM} & -6.09\% & -22.85\% & -13.88\% & -8.26\% & -58.48\% & -31.78\% \\
\texttt{JM-XGB} & -6.30\% & -17.46\% & -10.25\% & -6.79\% & -47.90\% & -21.62\% \\
\bottomrule
\end{tabular}

\caption{Performance metrics for the \mbox{0/1} strategy applied to individual assets over the testing period (2007-2023) using different regime forecasts. 
Top table shows Sharpe ratios, and bottom table shows maximum drawdown.
``\texttt{B\,\&\,H}'' represents the buy-and-hold strategy; 
``\texttt{JM}'' represents  the \mbox{0/1} strategy using regime forecasts solely from the jump model;
``\texttt{JM-XGB}'' represents the \mbox{0/1} strategy using regime forecasts from our hybrid framework.   %
The \mbox{0/1} strategy alternates between 100\% investment in the risky asset or 100\%  in the risk-free asset, depending on the forecasted regime. 
 A one-way transaction cost of 5 basis points is applied.
 Refer to Table \ref{tab:assets} for an overview of the asset universe.
}

    \label{tab:perf indiv assets}
\end{table}

Focusing on the maximum  drawdown (MDD) table, the \mbox{0/1} strategy consistently mitigates the downside risk across all assets; for instance, the MDD for \texttt{LargeCap} is more than halved. 
The strategy's ability to protect the downside underscores its practical relevance for risk management.
In terms of Sharpe ratios, the \texttt{JM} approach generally maintains or enhances these ratios compared to the buy-and-hold strategy. 
It's important to highlight the challenge of improving this risk-adjusted return measure due to the strategy's inherently reduced leverage. 
Under a simplistic assumption of independently and identically distributed returns with no transaction cost, randomly shifting investments to a risk-free asset to achieve a reduced leverage of $l < 1$ would proportionately reduce both volatility and returns by a factor of $\sqrt{l}$ and $l$, respectively, diminishing the Sharpe ratio by a factor of $\sqrt{l}$ , which is less than one. 
For instance, in the case of \texttt{Gold}, both the \texttt{JM} and \texttt{JM-XGB} strategies kept the investment out of the asset for over 60\% of the time, hence explaining our approaches' failure to  enhance Sharpe ratio for the asset.
The hybrid \texttt{JM-XGB} approach enhances Sharpe ratios across almost all assets, outperforming the simpler \texttt{JM} strategy, except in the Corporate Bond index. 
This improvement is expected due to our more sophisticated forecasting algorithm and the inclusion of additional macro-features. 
The degree of improvement varies, reflecting the differing predictability of each asset.

In summary, the financial significance of the regime forecasts generated by both the \texttt{JM} and \texttt{JM-XGB} approach is clear, highlighting the potential benefits of integrating these forecasts into practical portfolio management.
These advantages are consistent with previous empirical findings, such as those reported in \citet{shu2024regime, Bosancic2024}, which demonstrate the value of leveraging a regime-switching signal via the 0/1 strategy. 
While the comparison to the \texttt{JM} approach might not be fair enough -- given the lack of fine-tuning and the exclusion of macro-features in the \texttt{JM} approach -- the advantages of using a combined regime identification-forecasting framework becomes evident.
Therefore, the next subsection will focus exclusively on displaying the portfolio performance based on the \texttt{JM-XGB} approach.

\subsection{Asset Allocation Performance}   \label{subsec:portfolio res}

Following the demonstration of the financial relevance of our regime forecasts, we now examine dynamic asset allocation strategies that integrate these forecasts into diversified portfolios, through three portfolio models: MinVar, MV, and EW, with detailed methodologies presented in Section \ref{sec:portfolio}.

For comparison, We include a ``60/40'' fix-mix benchmark, with weights provided in \mbox{Table \ref{tab:60/40 weights}}.
This benchmark maintains an allocation of 40\% to the three bond indexes and 60\% to the remaining assets, rebalanced daily.
The allocation across different asset categories closely follows the benchmark weights used in \citet{Nystrup2018}.

\begin{table}[htbp]
    \centering
\begin{tabular}{*{9}{c}}
\toprule
\texttt{LargeCap} & \texttt{MidCap} & \texttt{SmallCap} & \texttt{EAFE} & \texttt{EM} & \texttt{REIT} & \texttt{HighYield} & \texttt{Commodity} & \texttt{Gold} \\
\midrule
 10.0\% & 5.0\% & 5.0\% & 5.0\% & 5.0\% & 10.0\% & 10.0\% & 5.0\% & 5.0\% \\
 \midrule
  \texttt{Treasury} & \texttt{Corporate} & \texttt{AggBond} &&&&&& \\
\midrule
 10.0\% & 10.0\% & 20.0\%   &&&&&& \\
\bottomrule
\end{tabular}
\caption{Benchmark 60/40 fix-mix portfolio weights, maintaining 60\% in equity, real estate, high-yield and commodity indexes (upper part of the table) and 40\% in three bond indexes (lower part of the table), rebalanced daily.}
    \label{tab:60/40 weights}
\end{table}

Table \ref{tab:port perf comp} presents a comprehensive comparison of seven asset allocation strategies: a 60/40 fix-mix portfolio, the original MinVar, MV, and EW portfolios, alongside their enhancements through our regime forecasts. 
The incorporation of regime forecasts evidently enhances all three models: the MinVar portfolio shows an improved return with reduced volatility, elevating the Sharpe ratio from 0.70 to 1.12, a pattern also seen in the EW portfolio. 
For the MV portfolio, which originally uses an exponentially smoothed historical mean that is too simplistic for robust performance, our regime forecasts yield a decent annualized excess return of 8.9\%, demonstrating once more the accuracy of regime forecasts.

\begin{table}[htbp]
    \centering
\begin{tabular}{l*{7}{r}}
\toprule
    & Fix-Mix  & \multicolumn{2}{c}{Minimum-Variance} & \multicolumn{2}{c}{Mean-Variance} &\multicolumn{2}{c}{Equally-Weighted} \\
    \cmidrule(lr){2-2} \cmidrule(lr){3-4} \cmidrule(lr){5-6} \cmidrule(lr){7-8}
    
 & 60/40 & MinVar &\makecell[l]{MinVar \\ (\texttt{JM-XGB})}  & MV & \makecell[l]{MV \\ (\texttt{JM-XGB})}  & EW & \makecell[l]{EW \\ (\texttt{JM-XGB})}  \\
\midrule
Return & 5.0\% & 2.8\% & 3.9\% & 2.6\% & 8.9\% & 5.5\% & 8.2\% \\
Volatility & 8.9\% & 4.0\% & 3.5\% & 7.1\% & 8.7\% & 10.8\% & 9.0\% \\
Sharpe & 0.57 & 0.70 & 1.12 & 0.37 & 1.02 & 0.51 & 0.91 \\
MDD & -31.5\% & -19.3\% & -7.1\% & -25.6\% & -13.5\% & -37.5\% & -17.6\% \\
Calmar & 0.16 & 0.15 & 0.55 & 0.10 & 0.66 & 0.15 & 0.47 \\
Turnover & 0.74 & 0.49 & 2.06 & 3.40 & 9.12 & 0.81 & 11.70 \\
Leverage & 1.00 & 1.00 & 0.91 & 0.95 & 0.86 & 1.00 & 0.92 \\
\bottomrule
\end{tabular}

    \caption{Performance table of seven asset allocation strategies over the testing period (2007-2023).
    Each row represents an annualized performance metric: excess return, excess volatility, Sharpe ratio, maximum drawdown, Calmar ratio, turnover and average leverage.
    ``60/40'' denotes the fix-mix strategy  shown in Table \ref{tab:60/40 weights};    %
    ``MinVar'', ``MV'', ``EW'' denotes the original minimum-variance, mean-variance and equally-weighted portfolios, respectively; 
    the other three columns with suffix ``\texttt{JM-XGB}'' denotes the respective portfolio model incorporating regime forecasts from our hybrid framework. 
    See Section \ref{sec:portfolio} for a detailed discussion of these methodologies.
    The annualized risk-free rate over the period is 1.1\%.
    A one-way transaction cost of 5 basis points is applied.
    }
    \label{tab:port perf comp}
\end{table}

Additionally, maximum drawdown is significantly mitigated across all models, most notably in the MinVar portfolio, where it is reduced from -19\% to -7\%, effectively creating an ``ultra-minimum-variance'' strategy. 
The Calmar ratio sees considerable improvement as well. 
The portfolio turnover, though increased by the discrete switches induced by our regime forecasts, remains at reasonable levels; MinVar (\texttt{JM-XGB}) records a turnover of 2.06 and MV (\texttt{JM-XGB}) 9.12, owing to the trading cost term in our optimizer's objective function. 
These enhanced portfolios markedly outperform the 60/40 fix-mix strategy, further validating the advantages of dynamic asset allocation.

Regarding the characteristics of different portfolio models,  the fix-mix strategy offers a moderate Sharpe ratio of 0.57 and an MDD of -31.5\%, offering itself as a stable but less optimized choice in portfolio management,  particularly in the absence of high-quality forecasts. 
The original MinVar portfolio already achieves a low volatility of 4.0\%, which further improves with \texttt{JM-XGB}. 
This portfolio type is desirable for risk-averse investors prioritizing stability over high returns and serves well as a defensive, low-beta diversifying component.
The traditional MV portfolio generally underperforms given a naive implementation, but significantly benefits from the integration of our regime forecasts, demonstrating its practical effectiveness given  satisfactory forecasting inputs. 
The EW portfolio inherently exhibits higher volatility and MDD due to its unfiltered exposure to risk.
Despite improved performance through \texttt{JM-XGB}, The EW portfolio faces challenges with high turnover not easily managed without formal optimization. 
It is crucial to note that the higher Sharpe ratio of 1.12 in the MinVar (\texttt{JM-XGB}) compared to 1.02 in the MV (\texttt{JM-XGB}) does not suggest a conclusion as to ``minimum-variance is superior to mean-variance in out-of-sample performance'' as claimed in \citet{demiguel2009optimal}; these models simply leverage our forecasting signals differently and exhibit unique risk-return characteristics.

To further investigate our enhancement of the MV portfolio, we calculate the correlation between our forecasted returns  and the actual returns, a crucial statistical criterion for measuring forecasting quality, as detailed in Table \ref{tab:corr comp}. 
Our \texttt{JM-XGB} methodology demonstrates significant improvements in forecasting accuracy across all assets, with especially notable gains in \texttt{EM} and \texttt{HighYield}, which show correlations of 6.02\% and 10.54\%, respectively. 
In particular, we gain positive correlation across all assets, with an overall forecasting correlation of 2.43\%.
In contrast, the simple EWMA estimates often result in negative correlations for many asset classes, indicating a reverse prediction of trends and highlighting the limitations of using a naive estimation based on historical returns.
This inefficacy of the naive estimates underlines the risk of careless implementation of MV optimization, which relies  on accurate return predictions to balance the trade-off between risk and return. %
The substantial enhancements in correlation with our \texttt{JM-XGB} approach suggest more reliable forecasting, providing a more robust foundation for informed asset allocation.

 \begin{table}[htbp]
    \centering
\begin{tabular}{l*{7}{r}}
\toprule
 & Overall & \texttt{LargeCap} & \texttt{MidCap} & \texttt{SmallCap} & \texttt{EAFE} & \texttt{EM} & \texttt{REIT} \\
\midrule
\texttt{EWMA} & -1.04\% & -1.58\% & -3.86\% & -3.72\% & -3.73\% & -2.03\% & -5.09\% \\
\texttt{JM-XGB} & 2.43\% & 1.66\% & 0.90\% & 1.03\% & 4.53\% & 6.02\% & 2.10\% \\
\midrule
 && \texttt{AggBond} & \texttt{Treasury} & \texttt{HighYield} & \texttt{Corporate} & \texttt{Commodity} & \texttt{Gold} \\
\midrule
\texttt{EWMA} && 1.25\% & -1.17\% & -0.16\% & -0.06\% & -1.05\% & -1.59\% \\
\texttt{JM-XGB} && 3.22\% & 1.64\% & 10.54\% & 2.62\% & 3.39\% & 0.32\% \\

\bottomrule
\end{tabular}

    \caption{Correlation between forecasted and actual asset returns over the testing period (2007-2023).
    ``\texttt{EWMA}'' denotes naive estimates based on an exponentially weighted moving average of historical returns with a 5-year halflife;
    ``\texttt{JM-XGB}'' denotes the return forecasts derived from our regime forecasts, as discussed in Section \ref{subsec:mv}.
    }
    \label{tab:corr comp}
\end{table}

\begin{figure}[htbp]
    \centering     %
    \includegraphics[height=.3\textheight]{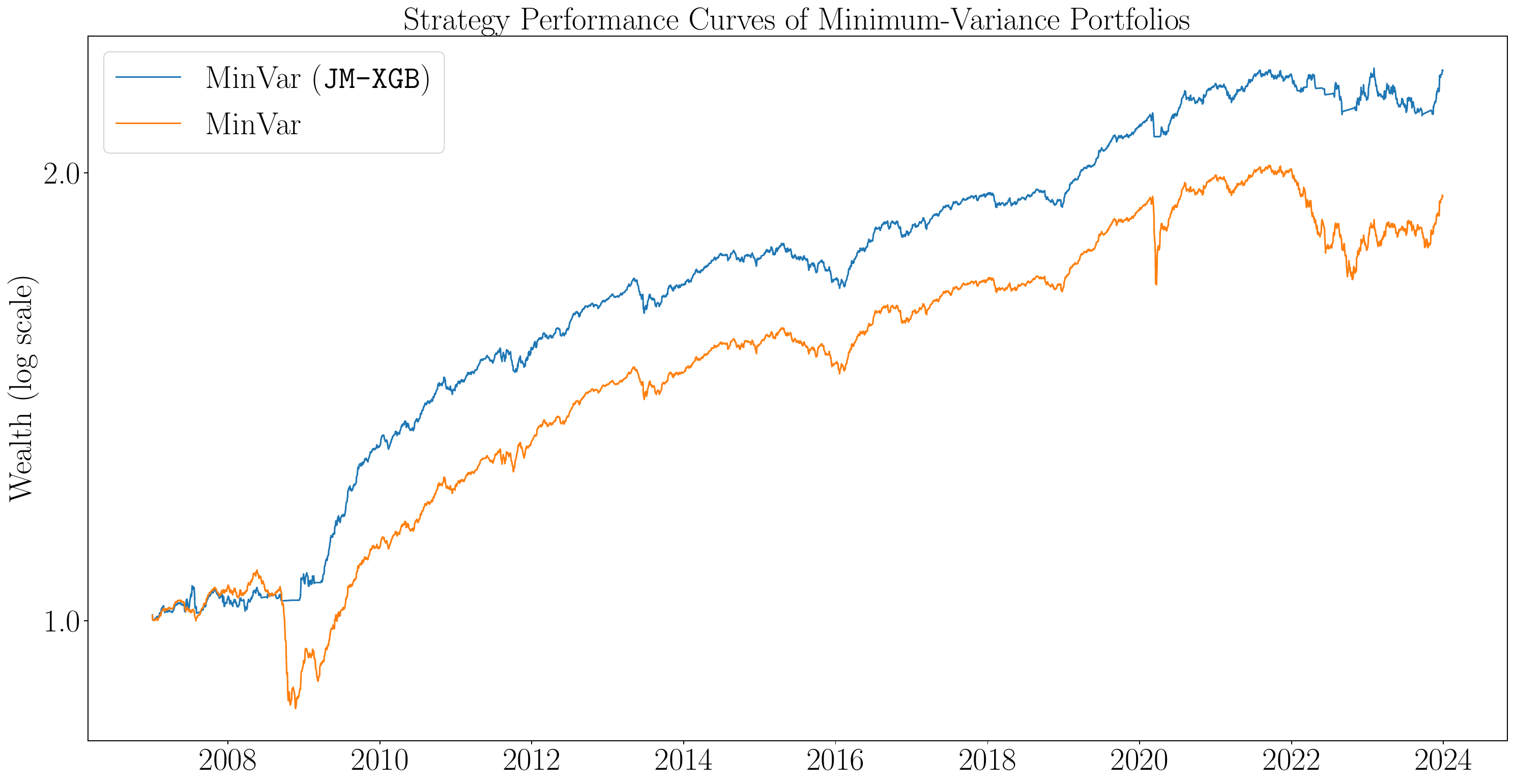}

    \includegraphics[height=.3\textheight]{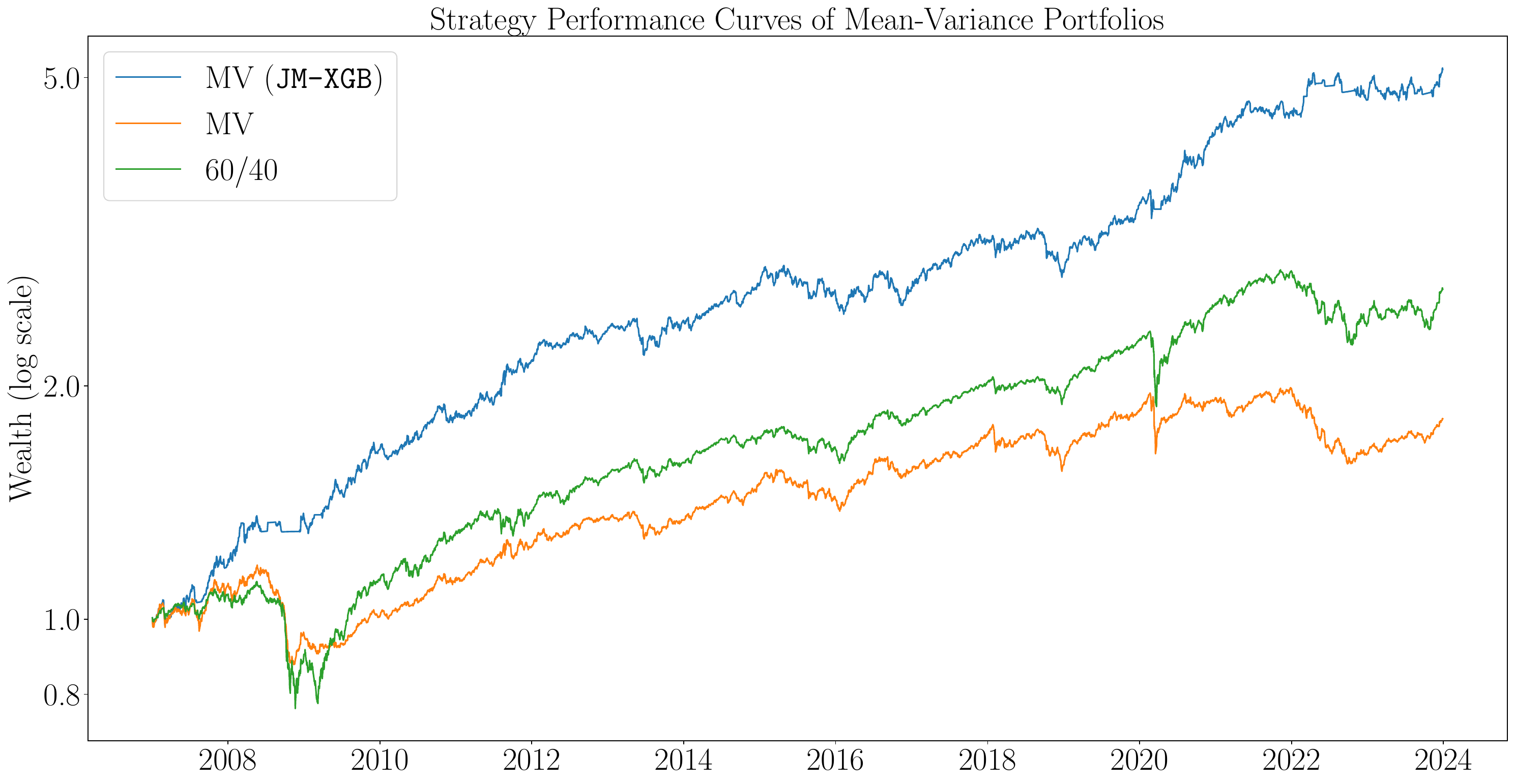}

    \includegraphics[height=.3\textheight]{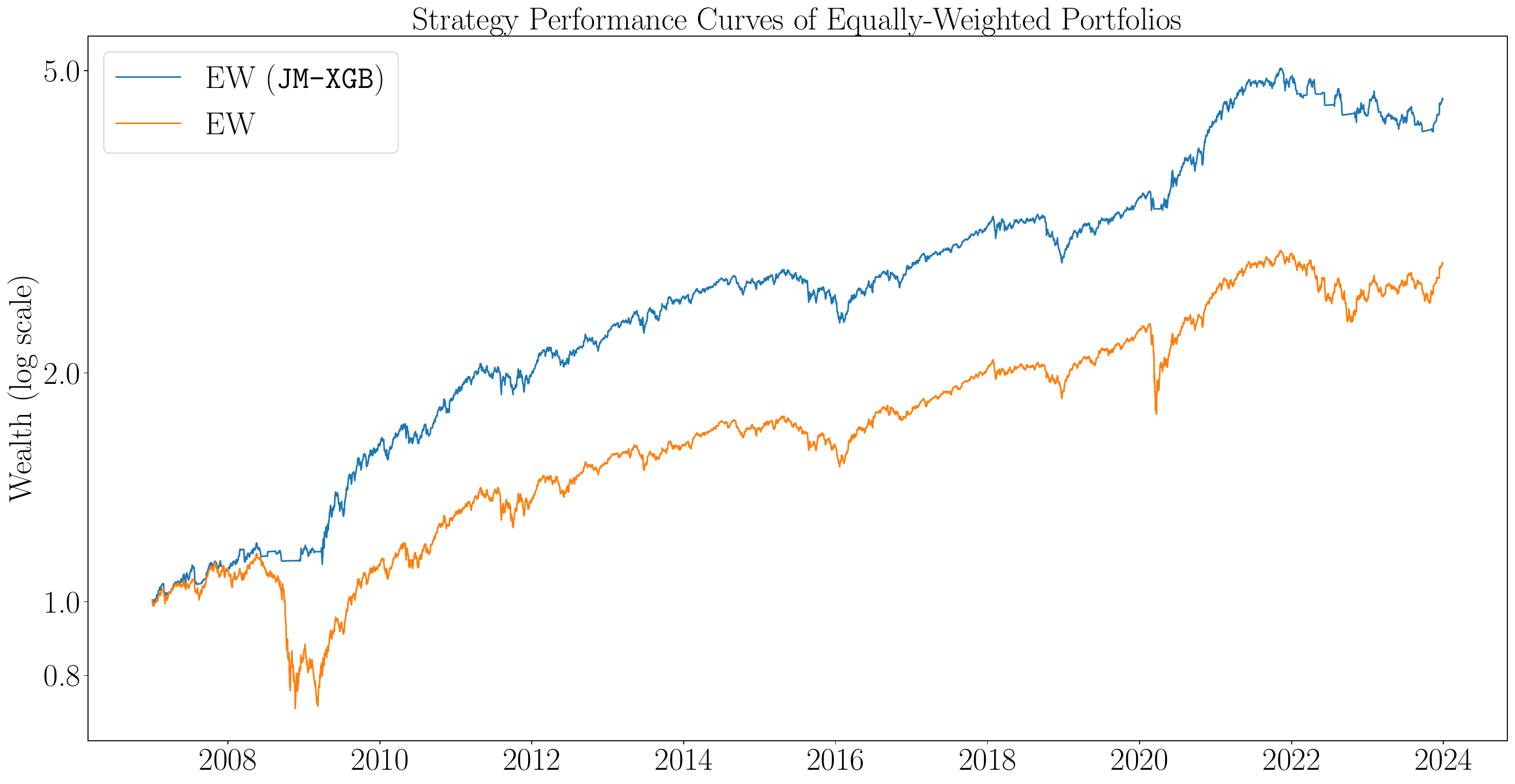}
    \caption{The wealth curves of different asset allocation strategies over the testing period (2007-2023), with minimum-variance, mean-variance, and equally-weighted portfolios displayed from top to bottom. 
     Yellow and blue curves represent the original portfolio model and the model incorporating our regime forecasts.
     ``60/40'' denotes the fix-mix strategy  shown in Table \ref{tab:60/40 weights}.   %
     A one-way transaction cost of 5 basis points is applied.
}
    \label{fig:perf curve all}
\end{figure}

To visualize the performance table, Figure \ref{fig:perf curve all} presents the wealth curves of the seven strategies. 
The graphs clearly demonstrate our outperformance, with robust downside protection during the global financial crisis, the COVID crash, and the 2022 market turmoil, showcasing the effectiveness of the regime forecasts during these persistent bearish markets.

We provide insights specifically for MinVar portfolios during the COVID crash and 2022 market turmoil, through portfolio weights.
During March 2020, the MinVar portfolio mainly held  40\% weights in \texttt{AggBond}, 36\% in \texttt{HighYield}, and 21\% in \texttt{Corporate}, with a leverage of 100\%, averaged across all trading days in this month. 
However, \texttt{HighYield} and \texttt{Corporate} suffered severe drawdowns of around -20\% and -15\%, respectively, over the month, leading to a heavy drawdown in the MinVar portfolio, which is undesirable for this presumably defensive strategy. 
In contrast, our MinVar (\texttt{JM-XGB}) portfolio held reduced average  weights of 14\% in \texttt{AggBond}, 13\% in \texttt{HighYield}, and 6\% in \texttt{Corporate}, with an average leverage of only 35\%, effectively avoiding the market downturn caused by the COVID shock. 
This protection is attributed to both our accurate regime forecasts and our strategic shift to the risk-free asset when only a limited number of assets are predicted to be bullish.

Similarly, in 2022, the poor performance of most bond indexes caused by surging interest rates led to the unsatisfactory performance of risk-based strategies such as MinVar and the risk-parity portfolio, which favor bonds heavily. 
Indeed, MinVar held average weights of 40\% in \texttt{AggBond}, 39\% in \texttt{HighYield}, and 13\% in \texttt{Corporate} with full leverage, which ends up with losses of -15\%, -13\%, and -17\% over the year, respectively, showcasing the slow reaction of covariance estimates to the high interest rate environment. 
In contrast, our MinVar (\texttt{JM-XGB}) portfolio held average weights of 14\% in \texttt{HighYield}, 8\% in \texttt{Corporate}, 7\% in \texttt{Gold}, and 5\% in \texttt{EAFE}, among other smaller weights, with an average leverage of 43\%, maintaining a flat portfolio performance despite the decline in almost all assets.

\subsection{Sensitivity Analysis}   \label{subsec:sense}

We conduct a basic sensitivity analysis to evaluate the robustness of our empirical results  against changes in the risk aversion parameter $\gamma^{\text{risk}}$ and trade aversion parameter $\gamma^{\text{trade}}$ appearing in the optimizer. 
Recall that, for our implementation of MinVar (\texttt{JM-XGB}) and MV (\texttt{JM-XGB}) portfolios, the default values are $\gamma^{\text{risk}}=10.0$ and $\gamma^{\text{trade}}=1.0$.
For practical applications, we recommend a thorough tuning of all the trading parameters based on backtests, as guided by \citet{boyd2024markowitz}.

For the MinVar portfolio, which exhibits low sensitivity to changes in the risk aversion parameter as outlined in Section \ref{subsec:minvar}, we explore the impact of the trading cost term by comparing performance with that under $\gamma^{\text{trade}}=0.0$, i.e., when the trading cost term is omitted from the objective function. 
The performance metrics for this analysis are presented in Table \ref{tab:sense minvar}.
The results show that even without considering transaction costs, the portfolio still performs robustly with a slightly decreased Sharpe ratio from 1.12 to 1.02, which remains significantly higher than that of the original MinVar portfolio.    %
Notably, the removal of the trading cost term  markedly increases turnover, from 2.06 to 11.80.
Therefore, using a non-zero trade aversion parameter  primarily serves to optimize risk-adjusted returns slightly while substantially reducing turnover to a practical level.

\begin{table}[htbp]
    \centering

\begin{tabular}{lrr}
\toprule
$\gamma^{\text{trade}}=$ & $0.0$ & $1.0$ (default) \\
\midrule
Return & 4.3\% & 3.9\% \\
Volatility & 4.2\% & 3.5\% \\
Sharpe & 1.02 & 1.12 \\
MDD & -12.8\% & -7.1\% \\
Calmar & 0.34 & 0.55 \\
Turnover & 11.80 & 2.06 \\
Leverage & 0.91 & 0.91 \\
\bottomrule
\end{tabular}

    \caption{
    Sensitivity analysis of the MinVar (\texttt{JM-XGB}) portfolio to changes in the trade aversion parameter $\gamma^{\text{trade}}$. 
    Performance metrics are compared between models with ($\gamma^{\text{trade}}=1.0$, default)  and without  ($\gamma^{\text{trade}}=0.0$) the trading cost objective term. 
}
    \label{tab:sense minvar}
\end{table}

Next, we investigate the influence of the risk aversion parameter $\gamma^{\text{risk}}$ on the MV (\texttt{JM-XGB}) portfolio by comparing $\gamma^{\text{risk}}$ values of 5.0 and 20.0, with our default value of 10.0, while keeping the trade aversion parameter fixed at 1.0, as shown in Table \ref{tab:sense mv}. 
Classical portfolio theory suggests that with a risk-free asset included, MV optimization always outputs the tangency portfolio across varying levels of risk aversion, and adjusts the allocation between the risk-free asset and the tangency portfolio accordingly, maintaining the Sharpe ratio at its maximal level.
Consistent with this theory, our analysis indicates that the Sharpe ratio remains stable at around 1.0 across different $\gamma^{\text{risk}}$ values, with higher risk aversion leading to reduced return and volatility, and lower leverage. 
Thus, our findings confirm that the empirical results are robust to changes in the risk aversion parameter.

\begin{table}[htbp]
    \centering
\begin{tabular}{lrrr}
\toprule
$\gamma^{\text{risk}}=$ & 5.0 & 10.0 (default) & 20.0 \\
\midrule
Return & 10.1\% & 8.9\% & 6.7\% \\
Volatility & 10.0\% & 8.7\% & 6.9\% \\
Sharpe & 1.01 & 1.02 & 0.96 \\
MDD & -15.4\% & -13.5\% & -13.5\% \\
Calmar & 0.65 & 0.66 & 0.49 \\
Turnover & 10.03 & 9.12 & 7.67 \\
Leverage & 0.89 & 0.86 & 0.76 \\
\bottomrule
\end{tabular}

    \caption{
    Sensitivity analysis of the MV (\texttt{JM-XGB}) portfolio to changes in the risk aversion parameter $\gamma^{\text{risk}}$. 
    Performance metrics are compared between $\gamma^{\text{risk}}$ values of 5.0, 10.0 (default) and 20.0.
}
    \label{tab:sense mv}
\end{table}

\section{Conclusion}  \label{sec:conclusion}

In this article, we introduce a hybrid regime identification-forecasting framework tailored to derive asset-specific regime forecasts for multi-asset portfolio construction. 
Unlike conventional methods that primarily focus on  broader economic conditions, our methodology harnesses the strengths of both unsupervised and supervised learning techniques to generate specific forecasts for individual assets. 
This approach begins with a robust unsupervised statistical jump model to classify historical periods into bullish or bearish regimes based on features derived from the asset return series. 
Subsequently, these labels are utilized to train a supervised gradient-boosted decision tree classifier, enhancing prediction accuracy by integrating both asset-specific and cross-asset macro-features.
The ultimate asset allocation weights are determined by incorporating these forecasts into the inputs to a Markowitz mean-variance optimizer. 
By tailoring regime identification to specific market dynamics and employing a dedicated classifier for forecasting, our framework aims to enhance asset allocation performance through more effective regime-switching strategies.

Our empirical study spanning from 1991 to 2023 across twelve diverse assets -- including global equity, bond, real estate, and commodity indexes -- illustrates the robustness and efficacy of integrating regime forecasts into dynamic asset allocation. 
Through illustrative figures, we visually display our method's capability to discern distinct market regimes for various assets.
Furthermore, we substantiate the financial benefits of our accurate regime forecasts for all assets, and observe marked improvements in portfolio performance across minimum-variance, mean-variance, and equally-weighted portfolio construction  by employing our novel framework, compared to conventional approaches.

Further extensions of our methodology could include expanding the regime identification-forecasting framework to encompass broader asset concepts such as adding a collection of long-short strategies to improve diversification benefits, refining feature selection tailored to asset-specific and time-varying dynamics for the regime identification component, improving the treatment of trading costs in portfolio models by reference to multi-period models, among possible enhancements.

\newpage

\appendix

\section{Dataset Details} \label{sec: data details}
Table \ref{tab:data details} provides a detailed description of the index names, with their Bloomberg tickers in brackets, and the tickers of actively traded index-tracking ETFs, used in our study.

\begin{table}[htbp]
\centering

\begin{tabular}{ccc}
\toprule
\textbf{Asset} & \textbf{Index Name (Bloomberg Ticker)} & \textbf{ETF Ticker} \\ 
\midrule
\texttt{LargeCap} & \makecell[c]{S\&P 500 Total Return Index \\ (\texttt{SPTR})} & \texttt{IVV} \\  
\midrule
\texttt{MidCap} & \makecell[c]{S\&P MidCap 400 Total Return Index \\ (\texttt{SPTRMDCP})} & \texttt{IJH} \\  
\midrule
\texttt{SmallCap} & \makecell[c]{Russell 2000 Total Return Index \\ (\texttt{RU20INTR})} & \texttt{IWM} \\
\midrule
\texttt{EAFE} & \makecell[c]{MSCI EAFE Net Total Return \\ USD Index (\texttt{NDDUEAFE})} & \texttt{EFA} \\
\midrule
\texttt{EM} & \makecell[c]{MSCI Emerging Net Total Return \\ USD Index (\texttt{NDUEEGF})} & \texttt{EEM} \\
\midrule
\texttt{AggBond} & \makecell[c]{Bloomberg US Aggregate Bond \\ Total Return Index (\texttt{LBUSTRUU})} & \texttt{AGG} \\
\midrule
\texttt{Treasury} & \makecell[c]{
Bloomberg US Long Treasury \\ Total Return Index (\texttt{LUTLTRUU})\tablefootnote{
Replaced by the S\&P U.S. Treasury Bond 20+ Year Total Return Index (\texttt{SPBDUSLT}) prior to 1994/3/2.
} 
} 
& \texttt{SPTL} \\
\midrule
\texttt{HighYield} & \makecell[c]{
iBoxx USD Liquid High Yield \\ Total Return Index (\texttt{IBOXHY})\tablefootnote{
Replaced by the total returns of Vanguard High-Yield Corporate Fund (\texttt{VWEHX}) prior to 1999/1/4.
}
} 
& \texttt{HYG} \\
\midrule
\texttt{Corporate} & \makecell[c]{Bloomberg US Corporate \\ Total Return Index (\texttt{LUACTRUU})} & \texttt{SPBO} \\
\midrule
\texttt{REIT} & \makecell[c]{
Dow Jones US Real Estate \\ Total Return Index (\texttt{DJUSRET})\tablefootnote{
Replaced by the Dow Jones Equity REIT Total Return Index (\texttt{REIT}) prior to 1992/1/3.
}
}
& \texttt{IYR} \\
\midrule
\texttt{Commodity} & \makecell[c]{Deutsche Bank DBIQ Optimum Yield Diversified \\ Commodity Index Excess Return (\texttt{DBLCDBCE})} & \texttt{DBC} \\
\midrule
\texttt{Gold} & \makecell[c]{LBMA Gold Price PM USD \\ (\texttt{GOLDLNPM})} & \texttt{GLD}\\
\bottomrule
\end{tabular}
\caption{Detailed overview of asset classes with corresponding index names (including Bloomberg tickers) and index-tracking ETF Tickers.}
\label{tab:data details}
\end{table}

\small
\bibliographystyle{apalike}
\bibliography{lit_regime}

\end{document}